\documentclass[aps,pra,aps,twocolumn,a4paper,superscriptaddress,longbibliography,nofootinbib,showpacs]{revtex4-1}
\usepackage[english]{babel}
\usepackage{graphicx}
\usepackage{amsmath}
\usepackage{amssymb}

\usepackage[colorlinks]{hyperref}

\usepackage[normalem]{ulem}   

\newcommand{\beq}{\begin{equation}}

\newcommand{\eeq}{\end{equation}}

\newcommand{\bra}[1]{\langle #1|}
\newcommand{\ket}[1]{|#1\rangle}

\newcommand{\comm}[2]{\left[#1,#2\right]}
\newcommand{\der}[2]{\frac{\partial #1}{\partial #2}}
\newcommand{\dsec}[2]{\frac{{\partial}^2 #1}{\partial {#2}^2}}

\newcommand{\ave}[1]{\langle #1 \rangle}

\begin{document}

\title{A Lindblad model of quantum Brownian motion}

\author{Aniello Lampo}\email{aniello.lampo@icfo.es}
        \affiliation{ICFO -- Institut de Ciencies Fotoniques, The Barcelona Institute of Science and Technology, 08860 Castelldefels (Barcelona), Spain}
\author{Soon Hoe Lim}
        \affiliation{Department of Mathematics, University of Arizona, Tucson, AZ 85721-0089, USA}
\author{Jan Wehr}
        \affiliation{Department of Mathematics, University of Arizona, Tucson, AZ 85721-0089, USA}
\author{Pietro Massignan}
        \affiliation{ICFO -- Institut de Ciencies Fotoniques, The Barcelona Institute of Science and Technology, 08860 Castelldefels (Barcelona), Spain}
\author{Maciej Lewenstein}
        \affiliation{ICFO -- Institut de Ciencies Fotoniques, The Barcelona Institute of Science and Technology, 08860 Castelldefels (Barcelona), Spain}
        \affiliation{ICREA -- Instituci{\'o} Catalana de Recerca i Estudis Avan\c{c}ats, Lluis Companys 23, E-08010 Barcelona, Spain}


\begin{abstract}
The theory of quantum Brownian motion describes the properties of a large class of open quantum systems.
Nonetheless, its description in terms of a Born-Markov master equation, widely used in literature, is known to violate the positivity of the density operator at very low temperatures.
We study an extension of existing models, leading to an equation in the Lindblad form, which is free of this problem.
We study the dynamics of the model, including the detailed properties of its stationary solution, for both constant and position-dependent coupling of the Brownian particle to the bath, focusing in particular on the correlations and the squeezing of the probability distribution induced by the environment. 
 \end{abstract}
\date{\today}

\maketitle
\section{Introduction}\label{Intro}
A physical system interacting with the environment is referred to as {\it open} \cite{BreuerBook}. 
In reality, such an interaction is unavoidable, so every physical system is affected by the presence of its surroundings, leading to dissipation, thermalization, and, in the quantum case, decoherence \cite{GardinerBook, SchlosshauerBook}. 

During the recent few years, interest in open quantum systems increased because of several reasons.
On the one hand, both decoherence and dissipation generally constitute the main obstacle to the realization of quantum computers and other quantum devices. 
On the other hand, recently there has been a series of very interesting proposals to exploit the interaction with the environment to dissipatively engineer challenging states of matter \cite{Griessner2006, Kraus2008, Diehl2010, Diehl2010b},
and of works where the engineering of environments paved the way to the creation of entanglement and superpositions of quantum states \cite{Poyatos1996, Turchette2000, Plenio2002}.

Moreover, open quantum system techniques are often adopted to investigate problems lying at the core of the foundations of quantum mechanics. 
Here, one of the unsolved issues regards the problem of the quantum-to-classical transition, i.e. the question how do classical features we experience in the macroscopic world arise from the underlying quantum phenomena \cite{SchlosshauerBook, Schlosshauer2005, Zurek2003, Haroche}.
Most of the theories addressing the emergence of the classical world deem it a consequence of the coupling of quantum systems with the environment 
 \cite{Maniscalco2006, Blume2008, Zurek2009, Korbicz2012, Maniscalco2014, Korbicz2014, Tuziemski2015}. 

In this work we focus on an ubiquitous model of open quantum system, the quantum Brownian motion (QBM), which describes the dynamics of a particle (playing the role of the open system) coupled with a thermal bath made up by a large number of bosonic oscillators (the environment) \cite{Caldeira1983a, Caldeira1983b, Haake1985, Grabert1988, Hu1992, Hu1993}.
QBM is in many situations the default choice for evaluating decoherence and dissipation processes, and in general it provides a way to treat quantitatively the effects experienced by an open system due to the interaction with the environment \cite{Marshall2003, Bose1997, Bose1999, Groblacher2015}. 
Hereafter, we will refer to the central particle studied in the model as \textit{the Brownian particle}.

The main tool for the investigation of the dynamics of the system is the master equation (ME), which describes the evolution of the reduced density matrix of the Brownian particle, obtained by taking the trace over the degrees of freedom of the bath.
The ME allows to compute various physical quantities, such as the time scales of decoherence and dissipation processes, as well as the average values of variables like position and momentum. 
 Widely used in the literature is the so-called Born-Markov master equation (BMME) \cite{SchlosshauerBook,BreuerBook}. The latter, however, does not always preserve  positivity of the density matrix, leading to violations of the Heisenberg uncertainty principle (HUP), i.e., $\sigma_X \sigma_P \geq \hbar/2$, especially at very low temperatures \cite{FlemPRE2011, Massignan2015}. Here $\sigma_X$ and $\sigma_P$ are the standard deviation of the position and momentum, respectively. The MEs in the so-called Lindblad form preserve the positivity of the density operator at all times \cite{Lindblad1976, SchlosshauerBook,BreuerBook}, and this in turn guarantees that the HUP is always satisfied. A brief, self-contained demonstration of the latter is given in the Appendix.
Various ways of addressing this difficulty have been put forward  \cite{Lindblad1976393, Diosi1993, Isar1994, Sandulescu1987, Gao1997, Wiseman1998, Gao1998, Ford1999, GaoReply, Vacchini2000}. In this paper we add a term to the BMME, that vanishes in the classical limit, bringing the equation to the Lindblad form and, in particular, ensuring that the HUP is always satisfied \cite{Lindblad1976}. We study the  dynamics of the obtained equation, in particular its stationary state.

An important purpose of the current paper is to investigate models of QBM more general than those usually studied in the literature. 
Usually, the particle-bath coupling is assumed to be linear in both the coordinates of the particle and the bath oscillators, and for definiteness in the following we will refer to this specific case with the name {\it linear QBM}.
We are here also interested in a more general case, where the coupling is still a linear function of the positions of the oscillators of the bath, but depends nonlinearly on the position of the Brownian particle.
This situation arises when dealing with inhomogeneous environments, in which damping and diffusion vary in space. 
An immediate application of this generalization concerns the physical behavior of an impurity embedded in an ultra cold gas.  In this case spatial inhomogeneities are due to the presence of trapping potentials and, possibly, stray fields \cite{Shashi2014, Tempere2009}.
Here, we will study in detail the case when the coupling depends quadratically on the position of the test particle, and we will refer to this case with the name {\it quadratic QBM}.

The paper is organized as follows.
In Sec.\ \ref{LinCase} we consider QBM with a linear coupling. 
We first introduce the BMME, and briefly discuss the lack of positivity preservation mentioned above. We then add a term to obtain a LME according to the procedure proposed in \cite{Gao1997}, and rewrite it in the Wigner function representation.  The Wigner function defines a quasi-probability distribution on the phase space \cite{GardinerBook}. We derive the time-dependent equations for the moments of this distribution, show that they have an exact Gaussian solution, and study in detail its long-time behavior.  In particular, we analyze the correlations induced by the environment, which cause a rotation and distortion of the distribution, as well as squeezing effects expressed by the widths and the area of the distribution's effective support.

In Sec.\ \ref{QuadraticCase} we study a non-linear QBM, corresponding to an inhomogeneous environment.
In particular we consider an interaction which is a quadratic function of the position of the Brownian particle. 
 This model has been studied in \cite{Massignan2015}, by means of a BMME.  
 We again modify the BMME to obtain an equation in the Lindblad form and we study its stationary solutions in the phase space (Wigner) representation.
For the quadratic QBM, the exact stationary state is no longer Gaussian but a Gaussian approximation can be used in certain regimes.  However, when the damping is strong, the Gaussian ansatz does not converge for large times, showing that it is not a good approximation to a stationary state.

\section{Linear QBM}\label{LinCase}

\subsection{The Model}
The QBM model describes the physical behavior of a particle interacting with a thermal bosonic bath of harmonic oscillators. In general the potential of the Brownian particle can be arbitrary,
but we will study the harmonic case only.
The model is described by the Hamiltonian:
\begin{equation}\label{Ham}
\hat{H}=\hat{H}_{S}+\hat{H}_{E}+\hat{H}_{I},
\end{equation}
in which:
\begin{align}
&\hat{H}_S=\frac{\hat{P}^2}{2m}+\frac{m\Omega^2\hat{X}^2}{2},\\\nonumber
&\hat{H}_E=\sum_k\frac{\hat{p}_k^2}{2m_k}+\frac{m_k\omega^2_k\hat{x}_k^2}{2},\\\nonumber
&\hat{H}_I=\sum_k g_k \hat{x}_k \hat{X},
\end{align}
where $m$ is the mass of the Brownian particle, $\Omega$ is the frequency of the harmonic potential trapping it, $m_k$ and $\omega_k$ are the mass and the frequency of the $k^{\rm th}$ oscillator of the environment, and $g_k$ are the bath-particle coupling constants.
In this Section, the interaction term $\hat{H}_I$ depends linearly on the positions of both the Brownian particle and the oscillators of the environment.  We refer to this model as a {\it linear QBM}.

The Hamiltonian is the starting point to derive the ME. 
We are interested in a BMME, obtained by making two approximations \cite{SchlosshauerBook}.
In the first one, the \textit{Born approximation}, we assume that the influence of the particle on the bath is negligible, so that the two systems remain uncorrelated (i.e. their joint state is a tensor product) at all times (including the initial one).
In the second, the \textit{Markov approximation}, we neglect memory effects, namely we require that the self-correlations created within the environment due to the interaction with the Brownian particle decay over a time scale much shorter than the relaxation time scale of the particle. 

Under these hypotheses, one derives from the Hamiltonian Eq.\ (\ref{Ham}) the following ME \cite{BreuerBook, SchlosshauerBook, Massignan2015},
\begin{align}\label{MEQBMLin}
\frac{\partial\hat{\rho}(t)}{\partial t}=&-\frac{i}{\hbar}\left[\hat{H}_S+C_x \hat{X}^2,\hat{\rho}(t)\right]\\
&-\frac{D_x}{\hbar}[\hat{X},[\hat{X},\hat{\rho}(t)]]
-\frac{D_p}{\hbar m\Omega}[\hat{X},[\hat{P},\hat{\rho}(t)]]\nonumber\\
&-\frac{iC_p}{\hbar m\Omega} [\hat{X},\{\hat{P},\hat{\rho}(t)\}].\nonumber
\end{align}
To avoid ambiguities, we wish to stress that, throughout the whole paper, we will be working in the Schr\"odinger picture, where the time-dependence is carried by the state of the system (rather than by the operators), and average values of operators are calculated as usual as $\ave{A}_t\equiv{\rm Tr}[\rho(t) A]$.

The effects of the bath on the motion of the Brownian particle are encoded in the spectral density of the bath. In the following, we will focus on the commonly used \textit{Lorentz-Drude} spectral density:
\begin{equation}\label{LDSD}
J(\omega)=\frac{m\gamma}{\pi}\frac{\omega}{1+\omega^2/\Lambda^2}
\end{equation}
which is linear at low frequencies, and decays as $1/\omega$ beyond the cut-off $\Lambda$, introduced to regularize the theory. With this choice of the spectral density the coefficients of the BMME at a bath's temperature $T$ read:
\begin{align}
\label{CxCoeff}
C_p&=\frac{m\gamma\Omega}{2}\frac{\Lambda^2}{\Omega^2+\Lambda^2}\\
C_x&=-\frac{\Lambda}{\Omega}C_p\\
D_x&=C_p\coth\left(\frac{\hbar\Omega}{2k_B T}\right)\\
D_p&
=\frac{2C_p}{\pi}\left[\frac{\pi k_B T}{\hbar \Lambda}+
z(T,\Lambda,\Omega)\right],
\end{align}
with: 
\begin{equation}
z(T,\Lambda,\Omega)=\psi\left(\frac{\hbar\Lambda}{2\pi k_B T}\right)-{\rm Re}\left[\psi\left(\frac{i\hbar\Omega}{2\pi k_BT}\right)\right]
\end{equation}
where 
$\psi(x)$ is the DiGamma function \cite{Massignan2015}.
In the following, we will refer to $T$, $\Lambda$, and $\gamma$ as the model's parameters. 
The term  proportional to $C_x$ in Eq.\ (\ref{MEQBMLin}) leads to a renormalization of the harmonic trapping frequency. Following the usual approach \cite{BreuerBook}, the latter may be cancelled by including  in the Hamiltonian the counter-term $\hat{V}_c=-C_x\hat{X}^2$. 

The evolution defined by Eq.\ (\ref{MEQBMLin}) does not preserve positivity of the density matrix. 
As discussed in detail in, e.g., Ref.\ \cite{Massignan2015, FlemPRE2011}, the lack of positivity leads to violations of the Heisenberg uncertainty principle away from the Caldeira-Leggett limit discussed below.  In particular, this prevents the study of the dynamics in the regime of very low temperatures.
 In fact, these violations in Eq.\ (\ref{MEQBMLin}) are driven by the logarithmic divergence at low temperatures  of $D_{p}$ (which is itself proportional to $\gamma$, i.e. to $g^2_{k}$).

Overcoming this problem is a fundamental step towards a correct description of the dynamics of a Brownian particle. In this section we propose a modified ME for the linear QBM which has the Lindblad form and, consequently, preserves positivity of the density matrix.   
It is well known that the ME\ (\ref{MEQBMLin}) cannot be expressed in the Lindblad form \cite{BreuerBook,SchlosshauerBook}.  Our equation differs from it by two terms, one of which can be naturally absorbed into the system's Hamiltonian.  

Adopting a LME is not the only possible manner to deal with the violations of the Heisenberg uncertainty principle. 
From a formal point of view, the ME\ (\ref{MEQBMLin}) is the result of a perturbative expansion to the second order in the strength of the bath-particle coupling (actually, expanding to second order requires weaker assumptions than the Born and Markov ones; the resulting equation may still take into account some non-Markovian effects which vanish in the limit of large times \cite{BreuerBook}).
In \cite{FlemPRE2011} it has been shown that Heisenberg principle violations in the stationary state disappear if one performs a perturbative expansion beyond the second order in the coupling constant. 
Obviously, if the exact ME is used, violation of Heisenberg principle cannot occur in any parameter regime.

\subsection{Lindblad Master Equation}
A LME has the form:
\begin{align}\label{LindEq}
\der{\hat{\rho}}{t}=&-\frac{i}{\hbar}\comm{\hat{H}_{S}}{\hat{\rho}}
+\sum_{i,j}\kappa_{ij}
\left[\hat{A}_{i}\hat{\rho}\hat{A}^{\dagger}_{j}-\frac{1}{2}\{\hat{A}^{\dagger}_i\hat{A}_j,\hat{\rho}\}\right],
\end{align}
where $\hat{A}_i$ are called Lindblad operators and $(\kappa_{ij})$ is a positive-definite matrix.

Following the approach proposed in \cite{Gao1997} we will replace the BMME (\ref{MEQBMLin}), which cannot be brought to a Lindblad form, by an equation of the form Eq.\ (\ref{LindEq}) with a single Lindblad operator of the form
\begin{equation}\label{LindbladOP}
\hat{A}_{1}=\alpha\hat{X}+\beta\hat{P},\qquad \textrm{ with } \kappa_{11}=1.
\end{equation}
Substituting this operator into Eq.\ (\ref{LindEq}) we obtain:
\begin{align}
\der{\hat{\rho}}{t}=&-\frac{i}{\hbar}\comm{\hat{H}'_{S}}{\hat{\rho}}-i\frac{\Gamma}{\hbar}\comm{\hat{X}}{\{\hat{P},\rho\}}\label{MELindNew}\\
&-\frac{D_{XP}}{\hbar^2}\comm{\hat{X}}{\comm{\hat{P}}{\hat{\rho}}}-\frac{D_{PP}}{2\hbar^2}\comm{\hat{P}}{\comm{\hat{P}}{\hat{\rho}}}\nonumber\\
&-\frac{D_{XX}}{2\hbar^2}\comm{\hat{X}}{\comm{\hat{X}}{\hat{\rho}}},\nonumber
\end{align}

with:
\begin{equation}\label{ShiftedHam}
\hat{H}'_{S}=\hat{H}_S-\frac{\Gamma}{2}\{\hat{X},\hat{P}\}\equiv\hat{H}_S+\Delta\hat{H} 
\end{equation}
and: 
\begin{align}
D_{XX}=&\hbar^2|\alpha|^2, &&D_{XP}=\hbar^2 \rm{Re}\left(\alpha^{*}\beta\right),\label{CoefficientsGenerator}\\\nonumber
D_{PP}=&\hbar^2|\beta|^2, &&\Gamma=\hbar \rm{Im}\left(\alpha^{*}\beta\right).
\end{align}
One could obtain the same result employing two Lindblad operators, proportional to  $\hat{X}$ and $\hat{P}$ respectively.
Without loss of generality, we may take $\alpha$ to be a positive real number
since multiplying $\hat{A}_1$ by a phase factor does not change Eq.\ (\ref{LindEq}), and we will restrict ourselves to ${\rm Im}\beta > 0$, because, as seen from Eq.\ (\ref{CoefficientsGenerator}), $\alpha{\rm Im}(\beta)$ is the damping coefficient $\Gamma$, which must be positive. 

Eq.\ (\ref{MELindNew}) differs from Eq.\ (\ref{MEQBMLin}) just by two extra terms,

involving $D_{PP}$ and $\Delta\hat{H}$.  Equating the coefficients of the remaining terms with those of the analogous terms appearing in  Eq.\ (\ref{MEQBMLin}), 
one finds:
\begin{align}
&D_{XX}=2\hbar D_x, &&D_{XP}=\frac{\hbar D_p}{m\Omega},\\\nonumber
&\Gamma=\frac{C_p}{m\Omega}, &&D_{PP}=\frac{(\hbar\Gamma)^2+D^2_{XP}}{D_{XX}}.
\end{align}
In the Caldeira-Leggett (CL) limit $k_BT\gg\hbar\Lambda
 \gg\hbar\Omega$, these reduce to:
\begin{align}
&\Gamma\approx\gamma/2,\\
&D_{XX}\approx2 m\gamma k_B T,\nonumber\\ 
&D_{XP}\approx-\gamma \frac{k_B T}
{\Lambda},\nonumber\\
&D_{PP}\approx\frac{\gamma k_BT}{2m \Lambda^2}\nonumber
\end{align}
Following  \cite{SchlosshauerBook}, since the quantities represented by $P$ and $m\Omega X$ have generally the same order of magnitude, one can argue, as in Eq. (5.56) of \cite{SchlosshauerBook}, that the terms proportional to $D_{XP}$ and $D_{PP}$ are negligible in the CL limit, recovering the structure of the usual CL ME.

The operator $\Delta\hat{H}$ can be absorbed into the unitary part of the dynamics defined by Eq.\ (\ref{MELindNew}), so it can be eliminated by introducing a counter term into the system's Hamiltonian.  More generally, we will add to $\hat{H}_S$ a counter term 
\begin{equation}
\hat{H}_{C}=(r-1)\Delta\hat{H},
\end{equation}
which depends on a parameter $r\in\mathbb{R}$, leading to the modified Hamiltonian:
\begin{align}
\hat{H}'_{S}=&\hat{H}_{S}-(r\Gamma/2)\{\hat{X},\hat{P}\}\\\nonumber
=&\frac{(\hat{P}-mr \Gamma \hat{X})^2}{2m}+\frac{m(\Omega^2-r^2\Gamma^2)\hat{X}^2}{2}.
\end{align}

The effect of $r$ is twofold: it introduces a gauge transformation which shifts the canonical momentum $\hat{P}$, and it renormalizes the frequency of the harmonic potential.
In the rest of the section we shall study the dynamics defined by equation Eq.\ (\ref{MELindNew}),  first for general values of $r$ and then, for the discussion of the stationary state, focusing on the case $r=0$.
We stress that the introduction of a counter term in the Hamiltonian does not affect the Lindblad character of the LME in Eq.\ (\ref{MELindNew}), since it just enters in its unitary part. 

\subsection{Solution of the LME}
We are interested in studying the long-time dynamics of the Brownian particle.
In particular, we consider its representation in the phase space, employing the Wigner function representation \cite{GardinerBook}.
In terms of the Wigner function, Eq.\ (\ref{MELindNew}) becomes $\dot{W}=\mathcal{L} W$, with
\begin{align}\label{LMEWF}
\mathcal{L}W=&-\frac{P}{m}\der{W}{X}+m\Omega^2X\der{W}{P}\\
&+\Gamma\left[r\der{}{X}(XW)+(2-r)\der{}{P}(PW)\right]\nonumber\\
&+\frac{1}{2}\left[D_{XX}\dsec{W}{P}+D_{PP}\dsec{W}{X}\right]-D_{XP}\frac{\partial^2 W}{\partial X\partial P}\nonumber.
\end{align}

Equivalently, one can look at the equations for its moments: 
\begin{align}
\label{EqMotMomW}\der{\ave{X}_t}{t}&=\frac{\ave{P}_t}{m}-r\Gamma{\ave{X}_t}\\
\der{\ave{P}_t}{t}&=-m\Omega^2 \ave{X}_t-(2-r)\Gamma\ave{P}_t\nonumber\\
\der{\ave{X^2}_t}{t}&=-2r\Gamma\ave{X^2}_t +\frac{2\ave{XP}_t}{m}+D_{PP}\nonumber\\
\der{\ave{XP}_t}{t}&=-m\Omega^2 \ave{X^2}_t -2\Gamma \ave{XP}_t + \frac{\ave{P^2}_t}{m} - D_{XP}\nonumber\\
\der{\ave{P^2}_t}{t}&=-2m\Omega^2 \ave{XP}_t -(4-2r)\Gamma \langle P^2 \rangle_t + D_{XX},\nonumber
\end{align}
where the moments of the Wigner function are calculated as
\begin{equation}
\ave{f(X,P)}_t=\int^{\infty}_{-\infty}dX\int^{\infty}_{-\infty}dP \; f(X,P)W(X,P,t).
\end{equation}
These moments correspond to symmetric ordering of the quantum mechanical operators $\hat X$ and $\hat P$ \cite{SchBook}. In particular, note that the time-dependence is solely contained in the Wigner function, in agreement with the fact that we work in the Schr\"odinger picture.

The solutions for the first moments are:
\begin{align}
\ave{X}_t= & e^{-\Gamma t }\left[X_{0}\cos(\beta_{r} t)+x^0_r\sin(\beta_{r} t)\right],\\ \nonumber
\ave{P}_t= & e^{-\Gamma t }\left[P_{0}\cos(\beta_{r} t)-p^0_r \sin(\beta_{r} t)\right],
\end{align}
where:
\begin{align}
&x^0_r=\frac{m\Gamma X_{0}(1-r)+P_{0}}{m\beta_r}\\
&p^0_r=\frac{\Gamma P_{0}(1-r)+m X_{0}\Omega^2}{\beta_r}\nonumber 
\end{align}
with:
\begin{equation}
X_0\equiv\ave{X}_0,\quad P_0\equiv\ave{P}_0,
\end{equation}
and:
\begin{equation}
\beta_r\equiv\sqrt{\Omega^2-\Gamma^2(r-1)^2}.
\end{equation}
Similar solutions have been presented in \cite{Kumar2009, Sandulescu1987, Isar1994}.
Eqs.\ (\ref{EqMotMomW}) may alternatively be written in terms of the kinetic momentum $\ave{\tilde{P}}_t=\ave{P}_t-m r \Gamma\ave{X}_t$: 
\begin{align}
&\der{\ave{X}_t}{t}=\frac{\ave{\tilde P}_t}{m},\label{EqXPtilde}\\
&\der{\ave{\tilde P}_t}{t} = -m\left[\Omega^2 -r(r-2)\Gamma^2\right]\ave{X}_t  -2\Gamma\ave{\tilde P}_t,\nonumber 
\end{align}
or equivalently gathered in the compact form
\beq
\dsec{\ave{X}_t}{t}+2\Gamma\der{\ave{X}_t}{t}+\left[\Omega^2 -r(r-2)\Gamma^2\right]\ave{X}_t=0.
\eeq
which, of course, can be derived directly from the equations Eq.\ (\ref{EqMotMomW}).
For both $r=0$ and $r=2$ one obtains a damped oscillator with the original frequency of the harmonic trap, $\Omega$.
For other values of $r$ the frequency is renormalized, with the maximal renormalization corresponding to $r=1$.  

In Eqs.\ (\ref{EqMotMomW}) we see that $r$ introduces apparent damping in the position, as already noted in \cite{Wiseman1998}.
Because of this, in the following we will set $r=0$. 
The extra term proportional to $D_{PP}$, not present in the starting BMME, appears only in the equation for $\dot{\ave{X^2}}$, without affecting the other equations, and in particular those for the first moments, so that it may be interpreted as a \textit{position diffusion coefficient}. 

We wish now to focus on the stationary solution of Eq.\ \eqref{LMEWF}. The latter may be found by means of the following Gaussian ansatz:
\begin{equation}\label{RGWF}
W_{ST}=\zeta\exp\left[\frac{1}{2(\rho^2-1)}\left(\frac{X^2}{\sigma^2_X}+\frac{P^2}{\sigma^2_P}+\frac{2\rho XP}{\sigma_X\sigma_P}\right)\right],
\end{equation}
which is normalized to one taking:
\begin{equation}
\zeta\equiv\frac{1}{2\pi\sigma_X\sigma_P\sqrt{1-\rho^2}},\quad |\rho|\leq1,
\end{equation}   
with:
\begin{equation}
\sigma_X=\sqrt{\ave{X^2}},\quad\sigma_P=\sqrt{\ave{P^2}},\quad\rho=-\frac{\ave{XP}}{\sigma_X\sigma_P},
\end{equation}
and, in the remainder of this Section, the variances are computed using the time-independent Gaussian Ansatz in Eq.\ \eqref{RGWF} \cite{Weedbrook2012}.
Inserting the Gaussian ansatz in Eq.\ (\ref{RGWF}) into Eq.\ (\ref{LMEWF}) we find:
\begin{align}
&\sigma^2_{X}=\frac{D_{XX}-4m\Gamma D_{XP}+m^2(4\Gamma^2+\Omega^2)D_{PP}}{4m^2\Gamma\Omega^2}\label{PosUnc}\\\nonumber
&\sigma^2_{P}=\frac{D_{XX}+m^2\Omega^2D_{PP}}{4\Gamma}\\\nonumber
&\sigma_{P}\sigma_{X}\rho=mD_{PP}/2
\end{align}
We introduce the adimensional variables:
\begin{equation}
\delta_x=\sqrt{\frac{2m\Omega \sigma^2_X}{\hbar}},\quad\delta_p=\sqrt{\frac{2\sigma^2_P}{m\Omega\hbar}}
\end{equation}
With this parametrization, the Heisenberg inequality $\sigma_X \sigma_P \geq \hbar/2$ reads $\delta_x\delta_p\geq1$.

The Lindbladian character of Eq.\ (\ref{LMEWF}) guarantees that the second moments will satisfy the Heisenberg relation at all times. We furthermore note that the term with coefficient $D_{PP}$, i.e. the extra term induced by the Lindblad form of the ME, leads to a correlation between the two canonical variables.

\begin{figure}
\begin{center}
\includegraphics[width=0.9\columnwidth]{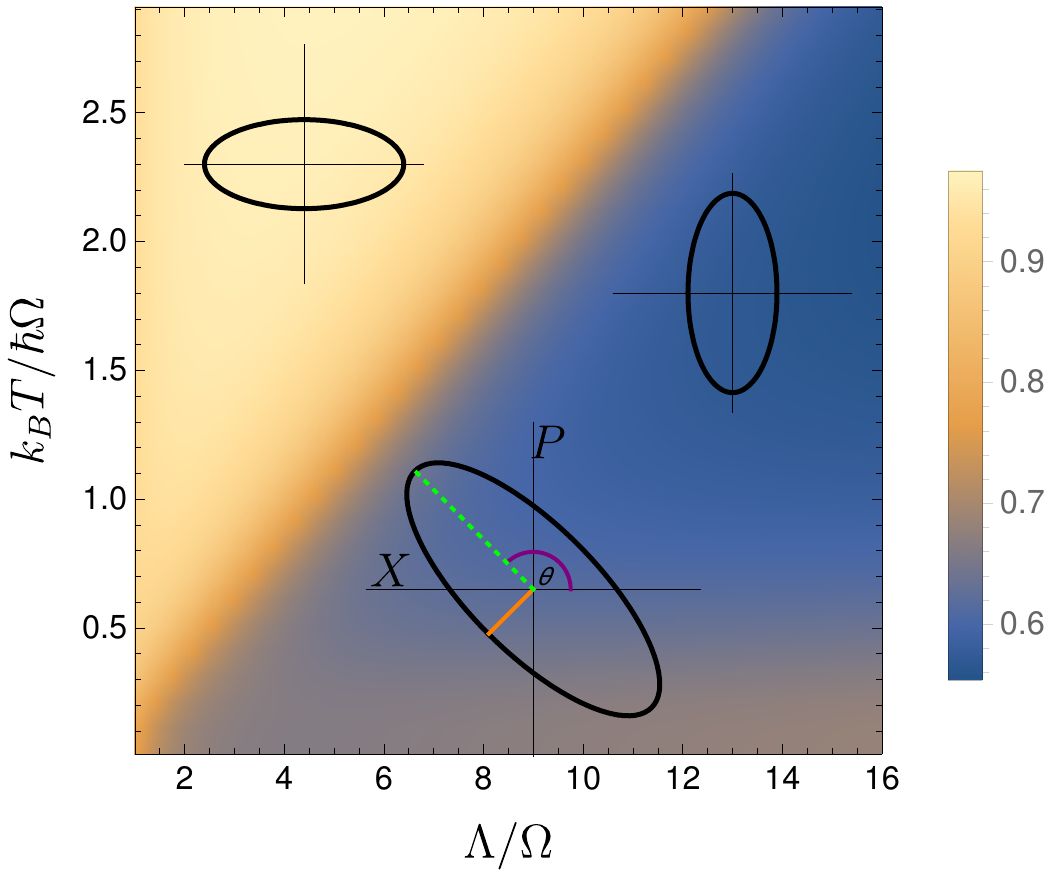}

\caption{\label{densityPlotAngle} Plot of the angle $\theta/\pi$ at $\gamma/\Omega=0.8$. 
This angle is represented in the ellipse at the bottom of the picture. Here, the orange-solid (green-dashed)  line represents the minor (major) axis of the Wigner function, i.e., that related to $\delta_l$ ($\delta_L$). The axes $X$ and $P$ are those of the phase space. 
}
\end{center}
\end{figure}

Geometrically, this correlation can be interpreted as a rotation of the stationary solution in the phase space, see the black sketches in Fig.\ \ref{densityPlotAngle}.
In the CL limit, the term with the coefficient $D_{PP}$ is negligible, and the solution is an ellipse with its axes parallel to the canonical ones, reproducing the well-known results.

To analyze the properties of the stationary state in the phase space, we consider the variances of the major and minor axes of the Wigner function. These axes are defined as the eigenvectors of the covariance matrix:
\begin{equation}
\text{cov}(X,P)=\left(\begin{array}{cc}
\delta^2_x&-\rho\delta_x\delta_p\\
-\rho\delta_x\delta_p&\delta^2_p\\
\end{array}\right)
\end{equation}
The smaller and larger eigenvalues of this matrix, $\delta_l$ and $\delta_L$, are given respectively by: 
\beq
\delta^2_{l,L}=\frac{1}{2}\left(\delta^2_x+\delta^2_p\mp\sqrt{\left(\delta^2_x-\delta^2_p\right)^2+4\delta^2_x\delta^2_p\rho^2}\right)
\eeq

We now aim to quantify such a rotation, calculating the angle $\theta$ between the major axis of the Wigner function (i.e. the eigenvector corresponding to $\delta_L$), and the $X$-axis of the phase space.
In Fig.\ \ref{densityPlotAngle} we present the behavior of $\theta$ as function of $T$ and $\Lambda$, at fixed $\gamma$.
At high $\Lambda$ the major axis aligns approximately with the $P$-axis of the phase space ($\theta = \pi/2$), 
while at low $\Lambda$, it is close to the $X$-axis ($\theta=\pi$), 
in agreement with the behavior of the BMME discussed in \cite{Massignan2015}, where $\ave{XP}$ was identically zero.
On the other hand, at low temperatures  the Wigner function associated to the stationary solution of the LME may be significantly rotated with respect to the axes of the phase space.

\begin{figure}
\centering
\includegraphics[width=0.9\columnwidth]{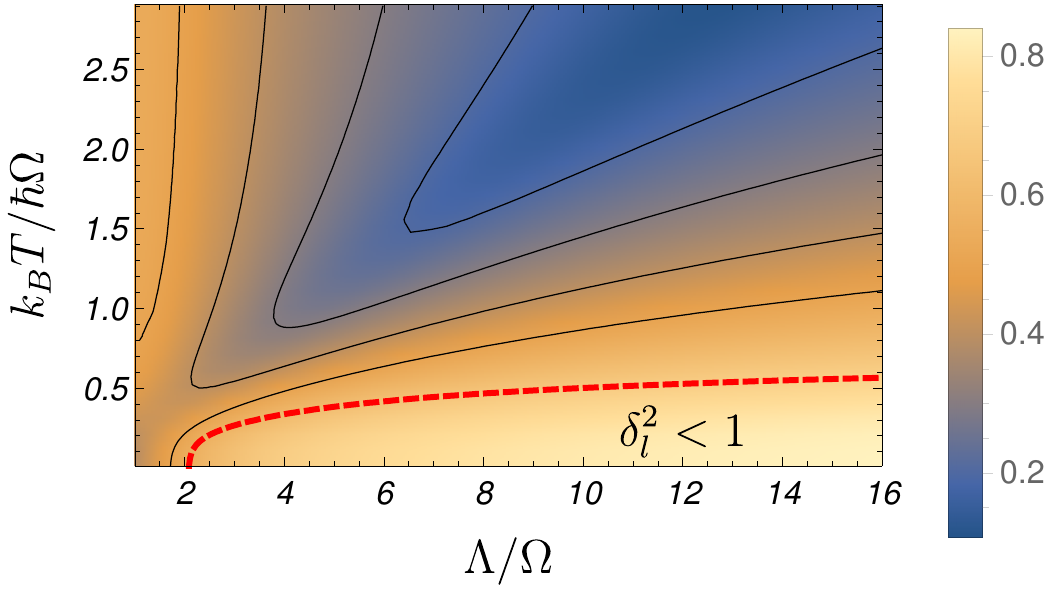}
\caption{Eccentricity of the Wigner function introduced in Eq.\ (\ref{RGWF}), at  $\gamma/\Omega = 0.8$.
The red dashed line represents the values of $T$ and $\Lambda$ yielding $\delta^2_l=1$, and we have genuine squeezing below it.
}
\label{sqWF}
\end{figure}

\begin{figure}
\centering
\includegraphics[width=0.9\columnwidth]{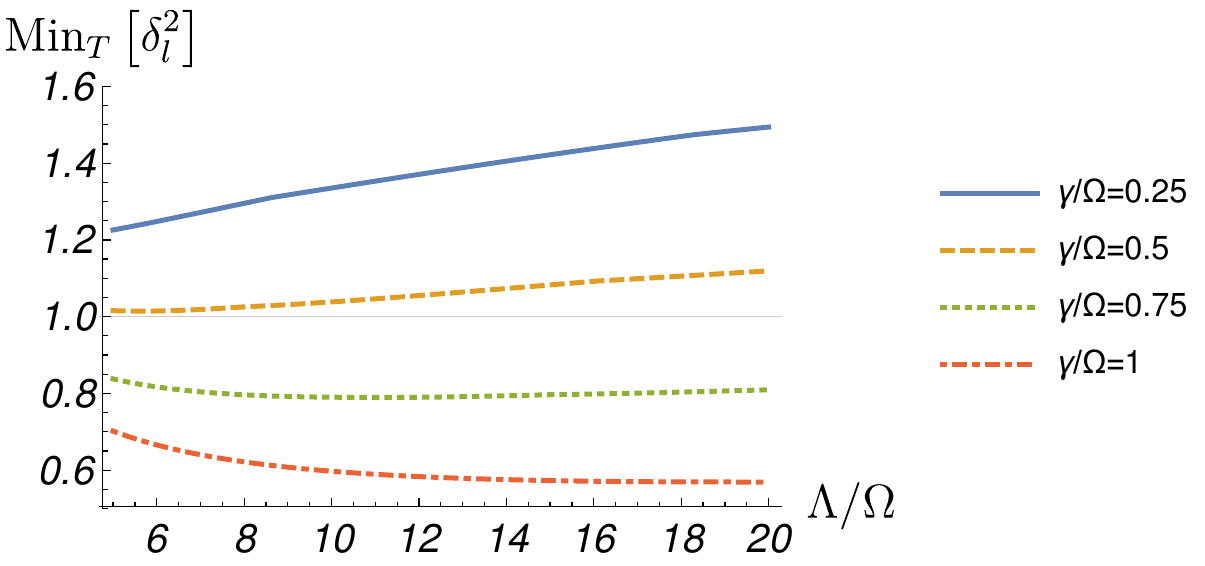}
\caption{Minimum value of $\delta^2_l$ over all temperatures, as a function of the cut-off frequency, at several values of the damping constant.}
\label{minimalSqueezing}
\end{figure}

In \cite{Massignan2015} it has been shown that, going to low temperature, the position of the Brownian particle governed by the BMME experiences \textit{genuine squeezing} along $x$ in the Wigner function representation, i.e. $\delta_x<1$. Similar squeezing effects are pointed out in \cite{Maniscalco2014}, by studying the numerical solution of the exact ME. In the case of the LME, it was checked numerically that $\delta_x$ introduced in Eq.\ (\ref{PosUnc}) is always bigger than one. However, the minor axis of the ellipse describing the Wigner function can display genuine squeezing.
To quantify the degree of squeezing of the Wigner function, 
Fig.\ \ref{sqWF} shows the values of eccentricity defined as
\begin{equation}
\eta=\sqrt{1-(\delta_l/\delta_L)^2},
\end{equation}
computed for different values of temperature $T$ and UV-cutoff $\Lambda$.
The eccentricity is largest at low temperatures. In particular, below the red dashed line, we find an area where $\delta_l<1$, corresponding to genuine squeezing along the minor axis of the Wigner Function, while in the CL limit the eccentricity $\eta$ approaches zero, and we obtain a Wigner function with circular symmetry.
In Fig.\ \ref{minimalSqueezing} we present the minimal value of $\delta^2_l$ obtained by choosing the appropriate (low) temperature. 
This picture highlights the range of values of $\Lambda$ and $\gamma$ where genuine squeezing occurs.  
We find that the eccentricity is an increasing function of the damping constant, i.e. squeezing becomes more pronounced as $\gamma$ grows.  In particular, at least $\gamma/\Omega>0.5$ is needed to obtain $\delta_l<1$.

\begin{figure}
\begin{center}
\includegraphics[width=0.9\columnwidth]{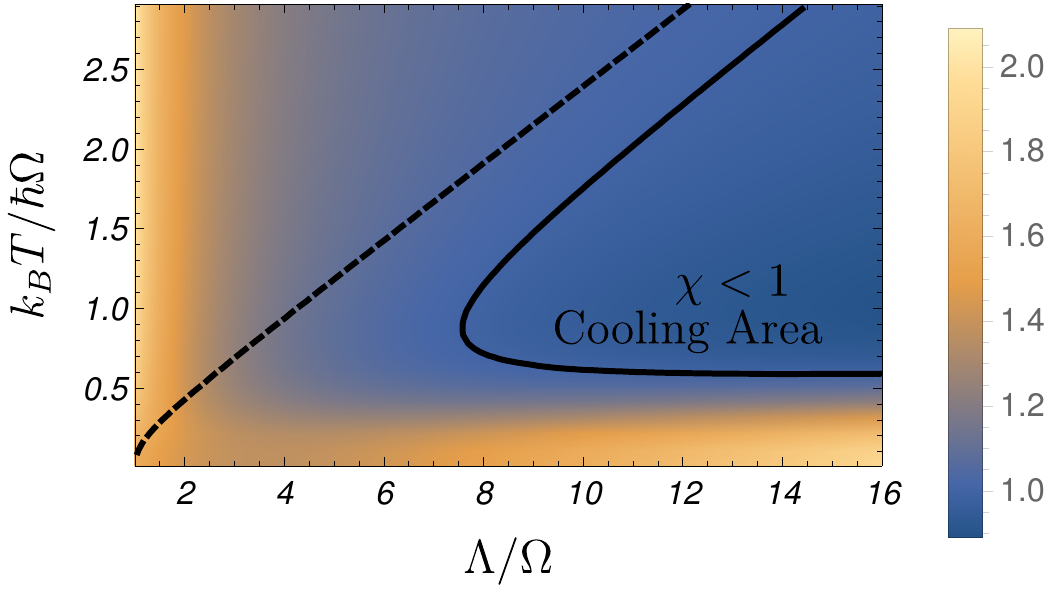}
\caption{\label{coolPlot} Cooling parameter $\chi$ introduced in Eq.\ (\ref{heatingCondition}), plotted for $\gamma/\Omega=0.8$. 
The system exhibits cooling to the right of the solid line, and heating to its left.
For comparison, the dashed line represents the cooling/heating boundary obtained with the BMME  (\ref{MEQBMLin}), which is independent of $\gamma$.
}
\end{center}
\end{figure}
 
\begin{figure}
\begin{center}
\includegraphics[width=0.9\columnwidth]{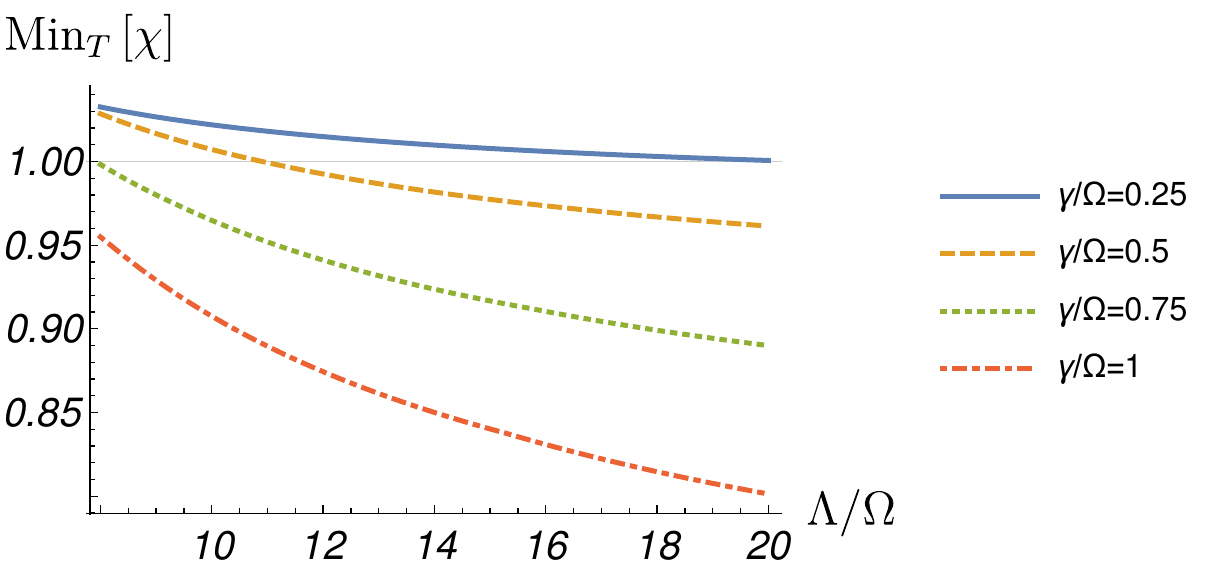}
\caption{\label{minimalCooling}
Minimum value of the cooling parameter $\chi$ over all temperatures, as a function of the cut-off frequency, at several values of the damping constant.
}
\end{center}
\end{figure}

We may say that the Brownian particle experiences an effective heating if the effective phase space area 
is larger than the one occupied by a quantum Gibbs-Boltzmann distribution at the same temperature. 
We thus define the system to be cooled if\footnote{For the Gibbs-Boltzmann distribution we have $\ave{X^2}_{GB}\ave{P^2}_{GB}\sim\coth^2{(\hbar\Omega/2k_BT)}$. So the denominator of Eq.\ (\ref{heatingCondition}) provides an information regarding the area of the Gibbs-Boltzmann distribution.}:
\begin{equation}
\chi=\frac{\delta_l\delta_L}{\coth\left(\frac{\hbar\Omega}{2k_BT}\right)}<1,
\label{heatingCondition}
\end{equation}
and heated otherwise.
The degree of heating/cooling $\chi$ is shown in Fig.\ \ref{coolPlot}.
In Fig.\ \ref{minimalCooling} we present the minimal value achieved by $\chi$ as the temperature is varied.
We note that to obtain small values of $\chi$ one needs to choose large values of both $\Lambda$ and $\gamma$.

There is a difference between the configuration of the cooling areas arising in the Lindblad dynamics studied here, and the ones produced by the BMME (\ref{MEQBMLin}) studied in \cite{Massignan2015}.
 In the latter, the cooling/heating boundary coincides with the line defined by $\delta_x=\delta_p$, and this condition does not depend on $\gamma$, while
in the present Lindblad model, the location of the boundary varies with  $\gamma$. 
However, the boundary calculated within the LME converges to the BMME one in the $\gamma\rightarrow0$ limit. Moreover, the LME discussed here displays heating at very low temperatures.

In Figs.\ \ref{minimalSqueezing} and \ref{minimalCooling} we have not extended the range of values of the damping constant beyond $\gamma=1$. 
In fact, the expressions for the coefficients of the equation Eq.\ (\ref{MELindNew}) have been obtained by comparing it with the equation Eq.\ (\ref{MEQBMLin}).  The latter is perturbative to second order in the strength of the coupling between the Brownian particle and the environment. 
The square of the coupling constant is proportional to the damping coefficient, so the validity of the perturbative expansion fails for $\gamma$ large. 
In particular, in the case of QBM this perturbative expansion holds for $\gamma\lesssim\Omega$  \cite{BreuerBook,Haake1985}.

\subsubsection*{Low Temperature Regime}
We consider here in detail the stationary state in the low temperature regime $k_BT<\hbar\Omega$.
Such a study was impossible in \cite{Massignan2015} because solutions violated the Heisenberg principle there. 
Here, the Lindblad form of the ME in Eq.\ (\ref{MELindNew}) ensures the positivity of the density matrix at all times, so no violations of the Heisenberg principle occur. 

In the discussion above, we noticed that the time-dependent equations of motion of the LME admit as an exact solution a Gaussian with non-zero correlations between the two canonical variables $X$ and $P$. In the stationary state, in particular, one finds $\ave{XP}=-mD_{PP}/2\neq0$. 
This is a novelty in comparison with the stationary solution of the BMME in Eq.\ (\ref{MEQBMLin}), which shows no correlations between $X$ and $P$. 
In the range of $\Lambda$ explored in Fig.\ \ref{densityPlotAngle}, the correlation between $X$ and $P$ becomes noticeable for $k_BT\lesssim0.5\hbar\Omega$. So, an important feature of the stationary solution of our LME at low temperature is that its major axis is rotated with respect to those of the phase space. 

In Fig.\ \ref{sqWF} we analyze the eccentricity of the stationary state. 
We point out that as the temperature decreases, the distribution becomes increasingly more squeezed. 
In particular, at low temperature we find a region displaying genuine squeezing of the probability distribution in the direction of $l$. 
 
 In Fig.\ \ref{coolPlot} we also note the presence of a cooling area in the low temperature regime. 
Nevertheless, in the zero-temperature limit the stationary state shows again heating.

The zero-temperature limit of the Lindblad model deserves special attention, as the two limits $T\rightarrow 0$ and $\gamma\rightarrow 0$ do not commute.
Taking first the zero-coupling and then the zero-temperature limit, one simply finds $\delta_x=\delta_p$ (in agreement with the general result for a free harmonic oscillator), but no further information on their specific value.
If instead one takes first $T\rightarrow 0$ and then $\gamma\rightarrow0$, one finds $\delta_x=\delta_p$ and the additional condition:
\begin{equation}\label{HeisenbergThreshold}
\delta_x\delta_p=\delta_l\delta_L=\frac{5}{4}+\frac{\left[\log(\Lambda/\Omega)\right]^2}{\pi^2}>1,
\end{equation}
indicating that for the Lindblad model the Heisenberg inequality is not saturated in the limit when the particle becomes free.
This is in contrast with the behavior of the non-Lindblad BMME (\ref{MEQBMLin}), for which, in this limit, we have $\delta_x\delta_p=1$. 
Summarizing, the effect of $D_{PP}$ is to introduce extra heating at low temperatures and couplings, manifested by a small constant, and a weak logarithmic dependence on the UV cut-off $\Lambda$.

\section{Quadratic QBM}\label{QuadraticCase}
\subsection{The Hamiltonian and the Lindblad ME}
In this section we consider the quadratic QBM, whose coupling is still linear in the positions of the oscillators of the bath, but is quadratic in the position of the Brownian particle: 
\begin{equation}\label{QuadraticCoupling}
\hat{H}_I=\sum_k \frac{g_k}{R} \hat{x}_k\hat{X}^2.
\end{equation}
Here $R$ is a characteristic length related to the motion of the Brownian particle and we set it to be 
$ R = \sqrt{\hbar/m \Omega}.$  The interaction term in Eq.\ (\ref{QuadraticCoupling}) describes an interaction of the particle with an inhomogeneous environment, giving rise to position-dependent damping and diffusion.

A concrete example where we may encounter this kind of nonlinearity is the model of an impurity in a Bose-Einstein condensate.
In \cite{Shashi2014, Tempere2009} it has been shown that the dynamics of such a system can be described by the \textit{Fr\"ohlich Hamiltonian}.
In an inhomogeneous gas, i.e. a gas with a spatially dependent density profile, this Hamiltonian differs from the QBM one due to the nonlinear dependence of the interaction term on the position of the impurity. 
When we consider, for instance, a Thomas-Fermi density profile, i.e. a density profile varying quadratically with the position, the interaction Hamiltonian is an even function of the position.  
Here, the coupling in Eq.\ (\ref{QuadraticCoupling}) provides the first-order correction to the zero-order term in the expansion of the interaction between the impurity and the bath. 
In short, QBM with a quadratic coupling is not just a mathematical exercise, but opens modeling possibilities in new contexts. In Appendix F of Ref.\  \cite{Massignan2015} it has been shown in detail that the Hamiltonian of an impurity in a BEC can be expressed in the form of that of QBM with a generic coupling.
  
The dynamics induced by the interaction term in Eq.\ (\ref{QuadraticCoupling}) has already been discussed in detail in \cite{Massignan2015}. There, the ME for the Brownian particle has been derived, in the BM approximations, for a Lorentz-Drude spectral density. Nevertheless, this ME is not in a Lindblad form, nor is exact. 
Accordingly,  the stationary solution is not defined for some values of the model's parameters  because of violations of the Heisenberg uncertainty principle at low temperatures.  

In this Section, we aim to find a LME as similar as possible to that derived in \cite{Massignan2015}.
Just like in the case of linear QBM, we expect it to differ from the BMME  by some extra terms. To achieve this goal we consider a single Lindblad operator:
\begin{equation}\label{LindOpQuadQBM}
\hat{A}_1=\mu\hat{X}^2+\nu\{\hat{X},\hat{P}\}+\epsilon\hat{P}^2
\end{equation}
where $\mu$, $\nu$ and $\epsilon$ are nonzero complex numbers.
Substituting it into Eq.\ (\ref{LindEq}) we obtain:
\begin{widetext}
\begin{align}\label{LMEwithquadraticcoupling}
\der{\hat{\rho}}{t}=&-\frac{i}{\hbar}\comm{\hat{H}_{S}+\Delta\hat{H}_{2}}{\hat{\rho}}-\frac{D_{\mu}}{2\hbar^2}\comm{\hat{X}^2}{\comm{\hat{X}^2}{\hat{\rho}}}
-\frac{D_{\nu}}{2\hbar^2}\comm{\{\hat{X},\hat{P}\}}{\comm{\{\hat{X},\hat{P}\}}{\hat{\rho}}}
-\frac{D_{\epsilon}}{2\hbar^2}\comm{\hat{P}^2}{\comm{\hat{P}^2}{\hat{\rho}}}\\
&-\frac{D_{\mu\nu}}{\hbar^2}\comm{\hat{X}^2}{\comm{\{\hat{X},\hat{P}\}}{\hat{\rho}}}
-\frac{D_{\mu\epsilon}}{\hbar^2}\comm{\hat{X}^2}{\comm{\hat{P}^2}{\hat{\rho}}}
-\frac{D_{\epsilon\nu}}{\hbar^2}\comm{\hat{P}^2}{\comm{\{\hat{X},\hat{P}\}}{\hat{\rho}}}\nonumber\\
&-i\frac{C_{\mu\nu}}{\hbar}\comm{\hat{X}^2}{\{\{\hat{X},\hat{P}\},\hat{\rho}\}}
-i\frac{C_{\mu\epsilon}}{\hbar}\comm{\hat{X}^2}{\{\hat{P}^2,\hat{\rho}\}}
-i\frac{C_{\epsilon\nu}}{\hbar}\comm{\hat{P}^2}{\{\{\hat{X},\hat{P}\},\hat{\rho}\}},\nonumber
\end{align}
\end{widetext}
where:
\beq
\frac{D_{\mu}}{\hbar^2}\equiv|\mu^2|,\quad\frac{D_{\mu\nu}}{\hbar^2}\equiv {\rm Re}(\mu^*\nu),\quad\frac{C_{\mu\nu}}{\hbar}\equiv {\rm Im}(\mu^*\nu),
\eeq
and similarly for the other combinations of indices. 
We could have obtained the same result by means of three Lindblad operators (rather than a single one), each proportional to one of the terms appearing on the right-hand side of Eq.\ (\ref{LindOpQuadQBM}).

Similarly to Sec.\ \ref{LinCase}, there is a term which appears in the unitary part of the ME:
\begin{align}
\Delta\hat{H}_2&=2D_{\mu\nu}\hat{X}^2-2D_{\epsilon\nu}\hat{P}^2+2D_{\mu\epsilon}\{\hat{X},\hat{P}\}\\
&-\frac{1}{2}C_{\mu\nu}\{\{\hat{X},\hat{P}\},\hat{X}^2\}-\frac{1}{2}C_{\mu\epsilon}\{\hat{P}^2,\hat{X}^2\}\nonumber\\
&+\frac{1}{2}C_{\epsilon\nu}\{\{\hat{X},\hat{P}\},\hat{P}^2\}.\nonumber
\end{align}
We eliminate it by introducing appropriate counter terms in the Hamiltonian. 

The ME in Eq.\ (\ref{LMEwithquadraticcoupling}) is in  a Lindblad form. 
Proceeding as in Sec.\ \ref{LinCase}, equating the coefficients on the right hand side of  Eq.\ (\ref{LMEwithquadraticcoupling})
to the corresponding ones in the BMME for quadratic QBM derived in \cite{Massignan2015}, we obtain:
\begin{align}
&D_{\mu\epsilon}=\frac{ D_{pp}}{m\Omega},\quad D_{\mu\nu}=D_{xp},\\
&C_{\mu\epsilon}=\frac{C_{pp}}{\hbar m\Omega},\quad C_{\mu\nu}=\frac{C_{xp}}{\hbar},\nonumber
\end{align}
and $D_{\mu}=2 m \Omega D_{xx}$. 
The remaining coefficients are then uniquely determined as: 
\begin{align}
&D_{\epsilon\nu}=\frac{1}{D_{\mu}}\left[D_{\mu\nu}D_{\mu\epsilon}+\hbar^2C_{\mu\nu}C_{\mu\epsilon}\right],\\\nonumber
&C_{\epsilon\nu}=\frac{1}{D_{\mu}}\left[C_{\mu\nu}D_{\mu\epsilon}-D_{\mu\nu}C_{\mu\epsilon}\right],\\\nonumber
&D_{\epsilon}=\frac{1}{D_{\mu}}\left[D^2_{\mu\epsilon}+\left(\hbar C_{\mu\epsilon}\right)^2\right],\\\nonumber
&D_{\nu}=\frac{1}{D_{\mu}}\left[D^2_{\mu\nu}+\left(\hbar C_{\mu\nu}\right)^2\right]. 
\end{align}
It is easy to check that in the CL limit $k_BT\gg\hbar\Lambda \gg\hbar\Omega$, the coefficients of all extra terms vanish,
and Eq.\ (\ref{LMEwithquadraticcoupling}) recovers the structure of the BMME introduced in Ref.\ \cite{Massignan2015}. \\

\subsection{Stationary State of the Quadratic QBM}\label{sec:StationaryStateQuadratic}
We turn now to the study of the stationary state of the Brownian particle in the case of  quadratic coupling. 
To this end we express the LME in Eq.\ (\ref{LMEwithquadraticcoupling}) in terms of the Wigner function $W$,
and obtain an equation of the form $\dot{W}=\mathcal{L} W$, with:
\begin{widetext}
\begin{align}\label{Wigner-quadratic}
\mathcal{L} =&-\frac{\partial_X
P}{m}+m\Omega^2\partial_P X
+2D_{\mu} \partial_P^2 X^2+2D_{\nu}\left(\partial_P P-\partial_X X\right)^2+2D_{\epsilon} \partial_X^2 P^2\\\nonumber
&+4D_{\mu\nu}(\partial_P^2 XP-\partial_P\partial_X X^2 +\partial_P X)-4 D_{\mu\epsilon} (\partial_X X -1) \partial_P P {-4 D_{\epsilon\nu}P\partial_X\left(\partial_P P-\partial_X X\right)}\\\nonumber
&+8C_{\mu\nu}\left[\partial_P P X^{2} +\frac{\hbar^2}{4}\partial_P^{2}(\partial_X X-1)\right]+C_{\mu\epsilon}\Big[4\partial_P XP^{2}-\hbar^2\partial_P\partial_X^{2}X+2\hbar^2\partial_P\partial_X\Big]\nonumber\\
&-2C_{\epsilon\nu}P\partial_X\left(4XP+\hbar^2\partial_P\partial_X\right). \nonumber
\end{align}
\end{widetext}

We now  find the stationary solution of the above equation. In this case the Gaussian ansatz in Eq.\ (\ref{RGWF}) may at best provide an approximate solution, in contrast with the case of the linear QBM, since the system of equations for the second moments is not closed. We approximate higher-order moments by their Wick expressions in terms of second moments (which would be exact in a Gaussian case), obtaining the following closed, nonlinear system of equations in the variables $\delta_x$, $\delta_p$ and $\rho$: 
\begin{widetext}
\begin{align}
\label{eqdelta1}\frac{1}{2}\der{\delta^2_{x}}{t}&=4m\hbar\Omega C_{\epsilon\nu}[1+\delta^2_x\delta^2_p(1+2\rho^2)]+2m^2\Omega^2D_{\epsilon}\delta^2_p+4D_{\nu}\delta^2_x
-\Omega\delta_x\delta_p\rho\\
\label{eqdelta2}\frac{1}{2}\der{\delta^2_{p}}{t}&=\frac{2D_{\mu}}{m^2\Omega^2}\delta^2_x-\frac{4\hbar}{m\Omega}C_{\mu\nu}+6\hbar C_{\mu\epsilon}\delta_x\delta^3_p\rho+\Omega\delta_x\delta_p\rho
+4\delta^2_p\left[D_\nu-D_{\mu\epsilon}-\frac{\hbar C_{\mu\nu}}{m\Omega}\left(1+2\rho^2\right)\delta^2_x\right],
\end{align}
and:
\begin{multline}
\label{eqdelta3}-\frac{1}{2}\der{(\delta_x\delta_p\rho)}{t}
=4\hbar C_{\mu\epsilon}+\Omega\delta^2_p-8m\Omega D_{\epsilon\nu}\delta^2_p+\frac{12\hbar}{m\Omega}C_{\mu\nu}\delta_p\delta^3_x\rho\\ 
+\left(8D_{\mu\epsilon}-12m\hbar\Omega C_{\epsilon\nu}\delta^2_p\right)\delta_x\delta_p\rho
-\left[\Omega+8\frac{D_{\mu\nu}}{m\Omega}+2\hbar\left(1+2\rho^2\right)C_{\mu\epsilon}\delta^2_p\right]\delta^2_x.
\end{multline}
\end{widetext}

This system of equations could admit more than one stationary solution,
 so we have to study the proper one. 
We choose the solution that coincides with that obtained with the non-Lindblad dynamics in the CL limit, since in this limit the coefficients of the extra terms of the LME in Eq.\ (\ref{LMEwithquadraticcoupling}) vanish. 
In \cite{Massignan2015} the stationary state in the case of the non-Lindblad dynamics has been studied in detail, and the variances have been calculated analytically.

\begin{figure}
\begin{center}
\includegraphics[width=0.9\columnwidth]{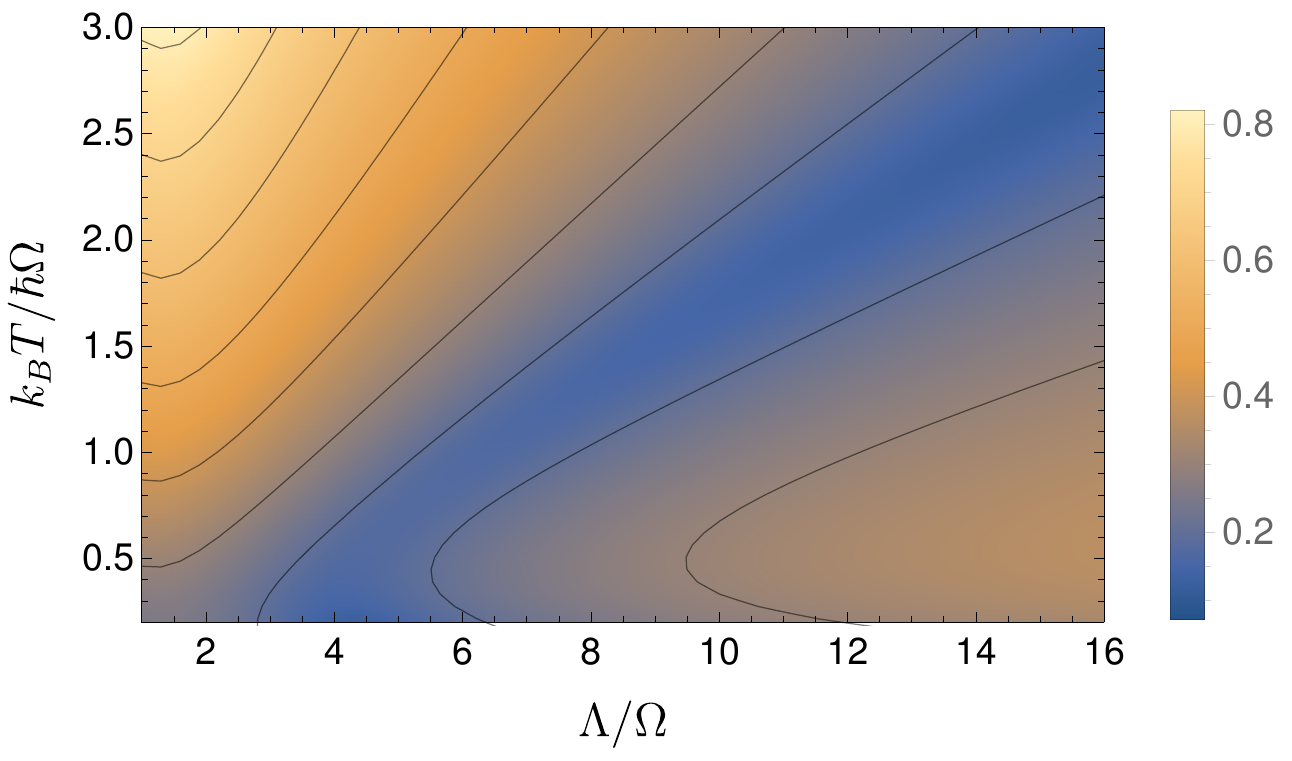}
\caption{\label{Eccentricity01} Eccentricity $\eta$ of the Wigner function at $\gamma/\Omega=0.1$, for quadratic coupling.  
}
\end{center}
\end{figure}

\begin{figure}
\begin{center}
\includegraphics[width=0.9\columnwidth]{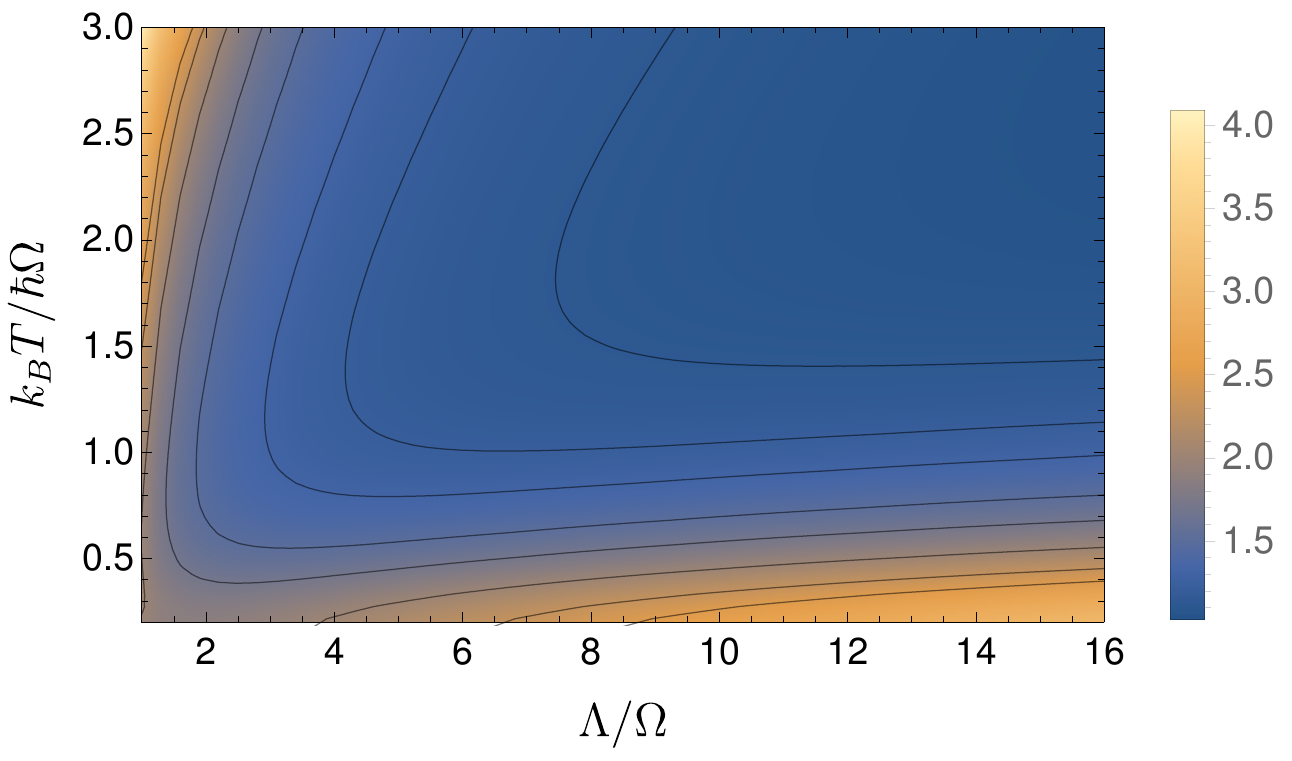}
\caption{\label{Cooling01} Cooling parameter $\chi$ for quadratic coupling, at $\gamma/\Omega=0.1$.  
}
\end{center}
\end{figure}

\begin{figure}
\begin{center}
\includegraphics[width=0.9\columnwidth]{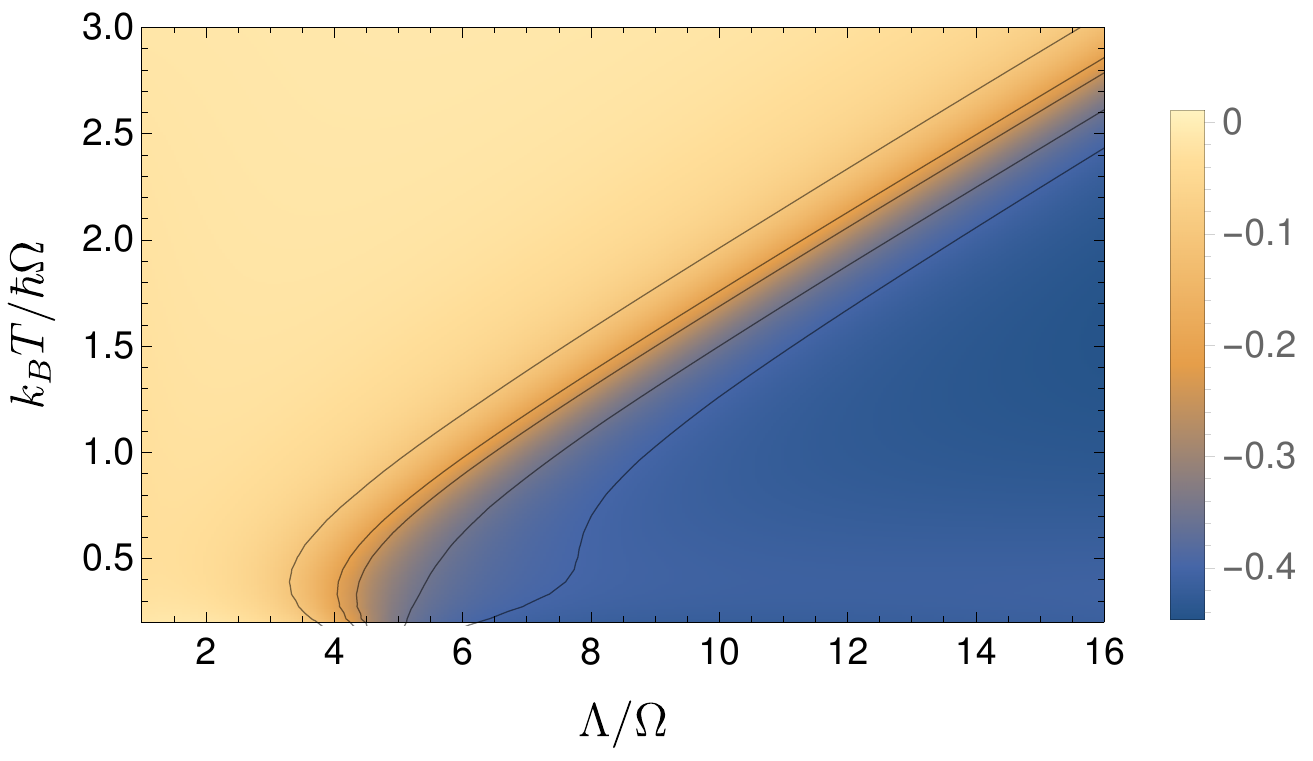}
\caption{\label{Angle01} Angle $\theta/\pi$ between the major axis of the Wigner function, and the X axis of the phase space at $\gamma/\Omega=0.1$, for quadratic coupling.  
}
\end{center}
\end{figure}

Similarly to the linear QBM studied in the previous section, we characterize the stationary state in terms of the variances of the Wigner function, and define the eccentricity, the cooling parameter, and the angle between the major axis and the X axis of the phase space as before. 
These quantities are shown  in Figs.\ \ref{Eccentricity01}, \ref{Cooling01}, and \ref{Angle01}, as functions of $\Lambda$ and $T$, when $\gamma/\Omega=0.1$. 
In Fig.\ \ref{Eccentricity01} we point out that the eccentricity tends to zero in the CL limit, while it increases away from it. This behavior is similar to that found for the linear QBM. 
We found that for $\gamma/\Omega\leq0.1$ the Brownian particle experiences neither cooling nor genuine squeezing. 

\begin{figure}
\begin{center}
\includegraphics[width=0.9\columnwidth]{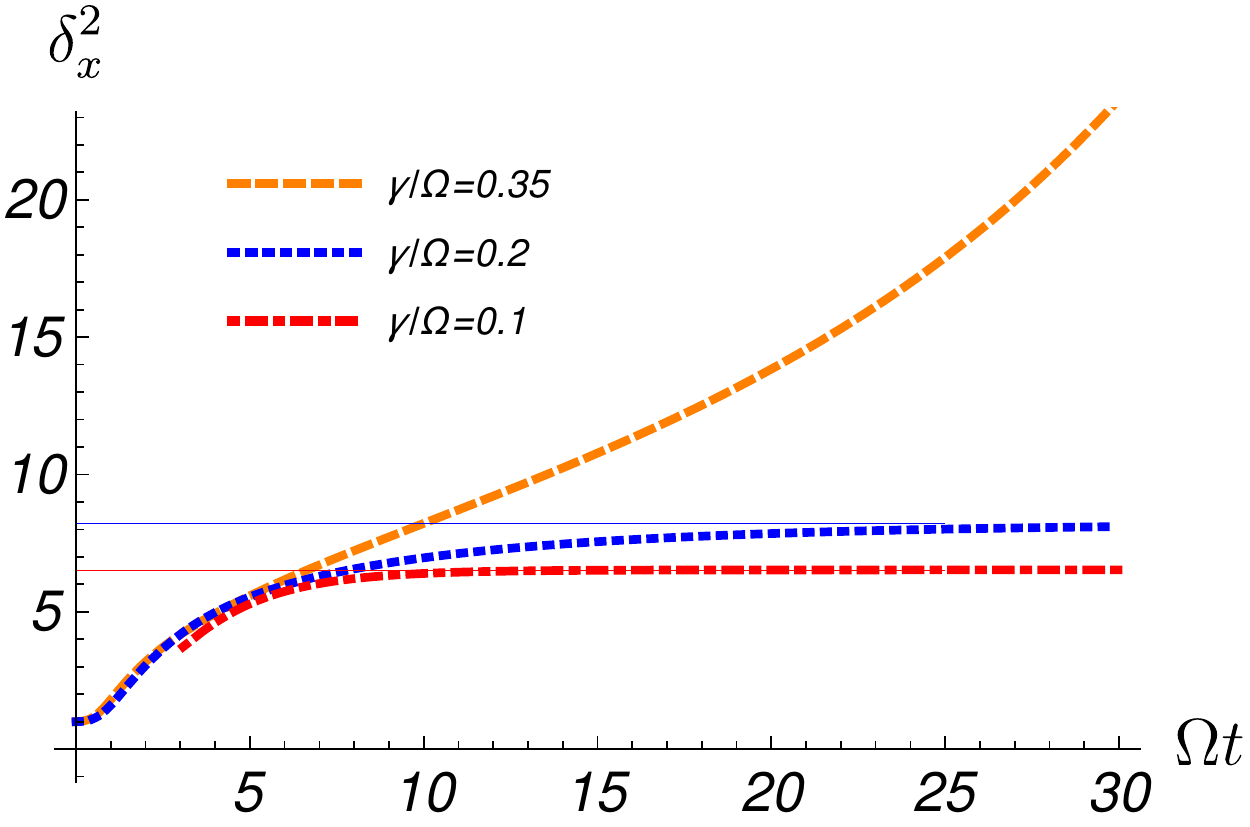}
\caption{\label{nostazstate} Time dependence of $\delta^2_x$ for several values of $\gamma$, at $\Lambda/\Omega=16$ and $k_B T/\hbar\Omega=4$. The thin solid lines represent the stationary value of $\delta^2_x$ in the state, namely the stationary solution of Eqs.\ (\ref{eqdelta1}-\ref{eqdelta3}) for such a quantity.   
}
\end{center}
\end{figure}

In contrast to the linear case, we do not find a noticeable rotation at low temperature in the quadratic one. We would expect to observe this at larger values of $\gamma$, as in the case of linear coupling.
 However, for larger values of the damping constant the many stationary solutions of the system of Eqs.\ (\ref{eqdelta1}-\ref{eqdelta3}) cross, and therefore it is not straightforward to determine the stationary solution of
 \eqref{Wigner-quadratic} that coincides with the one obtained in the CL limit. 
Moreover, for larger values of $\gamma$ the Gaussian ansatz given in  Eq.\ (\ref{RGWF})  may fail to approximate any stationary states. 
To show this point, in Fig.\ \ref{nostazstate} we plotted the time dependence of $\delta^2_x$ for several values of $\gamma$, at fixed values of $T$ and $\Lambda$.
Above a certain value of $\gamma$, the position variance does not converge to a stationary value. This suggests that in these cases the Gaussian solution of Eq.\ \eqref{Wigner-quadratic} is not stationary.   
Fig.\ \ref{nostazstate} is plotted for the initial conditions $\delta^2_x=\delta^2_p=1$, corresponding to the case when the  harmonic oscillator is in its ground state.  
The choice of the initial conditions is not crucial, as we observe a very similar behavior with quite different initial conditions.

\begin{figure}
\begin{center}
\includegraphics[width=0.9\columnwidth]{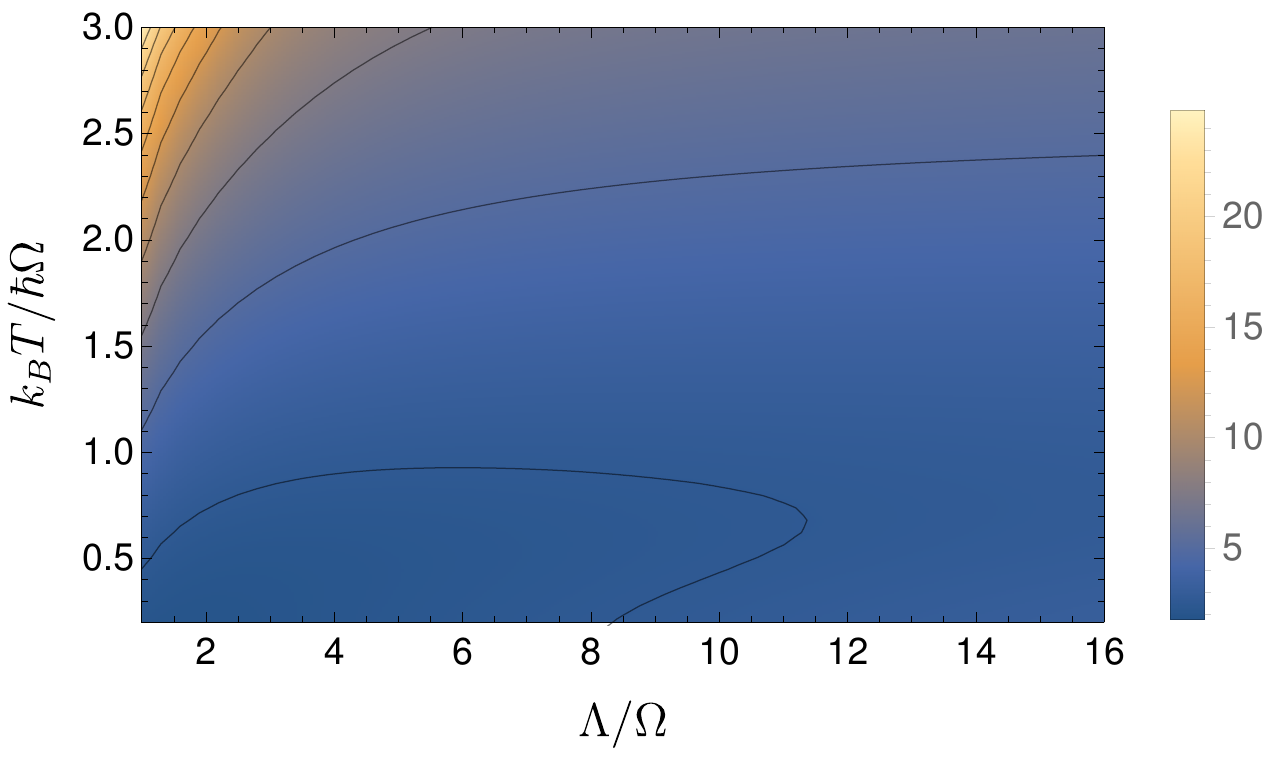}
\caption{\label{Heisenberg01} 
Plot of the product $\delta_x\delta_p$ at $\gamma/\Omega=0.1$, for quadratic coupling. This quantity is always larger than 1, in accord with the HUP.
}
\end{center}
\end{figure}

We conclude this Section pointing out that, although in Eqs.\ (\ref{eqdelta1}-\ref{eqdelta3}) we performed the Gaussian approximation at the level of the equations for the moments, it is possible to obtain exactly the same result applying the approximation directly on the original LME in Eq.\ (\ref{LMEwithquadraticcoupling}), or on that LME expressed in terms of the Wigner function,
Eq.\ (\ref{Wigner-quadratic}).
In Appendix \ref{appendixGauss} we show, by a very general analytical demonstration, that the Gaussian approximation applied to the original LME yields again a ME of the Lindblad form, guaranteeing therefore that the approximated solutions will preserve the HUP at all times.
We provide further numerical evidence of this fact in Fig.\ \ref{Heisenberg01}, where we plot the product of the two uncertainties $\delta_x$ and $\delta_p$ resulting by Eqs.\ (\ref{eqdelta1}-\ref{eqdelta3}), on which the Gaussian approximation has been carried out. As may be noticed in the figure, the approximated equations do not produce any violation of the HUP.

\section{Conclusions and Outlook}
We studied a modification of the QBM model, focusing on the description of the stationary state of the Brownian particle in the phase space, using the Wigner function representation.
To perform this analysis we considered a ME of the Lindblad form, which ensures the positivity of the density matrix at all times. 
In this way we got rid of the Heisenberg principle violations discussed in \cite{Massignan2015}, which prohibited the study of the dynamics in the low temperature regime. 

In Sec.\ \ref{LinCase} we dealt with QBM with a linear coupling. In this case, the stationary state can be represented exactly by a Gaussian Wigner function.  
We put particular emphasis on the analysis of its properties in the low temperature regime, where its properties are most interesting.

At low temperature we found that the Brownian particle exhibits genuine squeezing of the probability distribution.
An important feature of the stationary state in this regime is its rotation in the phase space, a direct consequence of the extra terms introduced to obtain a Lindblad form for the equation. 
Another important effect experienced by the stationary state can be quantified by the degree of cooling, expressing the ratio of the area of the effective support of the Wigner function to that of the Gibbs-Boltzmann distribution at the same temperature.
A concrete physical system where cooling and squeezing can be encountered is suggested in \cite{Maniscalco2004}. 
 
In Sec.\ \ref{QuadraticCase} we performed the same analysis for QBM with a coupling which is quadratic in the coordinates of the test particle. Importantly, we found that there exists a critical value of the damping constant over which the Gaussian ansatz fails to approximate any stationary solutions. 

Our procedure of adding extra terms to the BMME derived in \cite{Massignan2015}, so that the resulting equation is in a Lindblad form, is just one of the ways to obtain a Markovian dissipative LME. Other approaches have been presented, e.g., in Refs.\ \cite{Taj2008, Pepe2012}.
Moreover, for Gaussian dynamics an exact (non-Markovian) closed master equation with time dependent coefficients can be derived  \cite{Ferialdi2014,Ferialdi2016March,Carlesso16}. 
In a forthcoming work we plan to derive a LME describing QBM with a general class of couplings, and study its various limiting behaviors, in particular the small mass limit of the Brownian particle.  

The method which we used to treat the LME  in this manuscript is not the only suitable one. Another possible manner to solve this kind of equations, and in particular to characterize the stationary solution has been presented in Ref.\ \cite{Englert1994}. The core of this procedure is turning LMEs into partial first-order differential equations for a phase-space distribution (PSD) which generalizes well-known ones such as the Wigner function. The main point lies in removing the evolution generated by the free Hamiltonian by including it in the interaction representation. Accordingly, the time dependence of the PSD originates solely from the interaction term.
Although the interaction picture adopted in \cite{Englert1994} could be used in the context we are treating, its usefulness is not necessarily guaranteed. In fact, the interaction picture represents a suitable tool when the free part of the Hamiltonian describes a dynamics much faster than that induced by the interaction term. In general this is not the case for the Brownian motion of a trapped particle, where the time scales related to both processes can approach the same order of magnitude.   
On the other hand, employing this method looks like a very interesting task, which maybe can allow us to go beyond the Gaussian approximation underlying Sec. (\ref{QuadraticCase}). This task, however, lies outside of the focus of the current paper. We thus reserve it for future works.

There are other methods to correct the Heisenberg principle violations 
highlighted in \cite{Massignan2015}. The BMME for the quadratic QBM derived in \cite{Massignan2015} is based on a second order perturbative ME in the bath-particle coupling constant. Going to higher orders permits one, in principle,  to get rid of the violations of the Heisenberg principle. This task can be pursued by means of the time-convolutionless method presented in \cite{BreuerBook}. An advantage of this approach is that the resulting ME incorporates non-Markovian effects. 
Nevertheless, since it arises from a perturbative expansion, it does not allow to investigate the strong coupling regime $\gamma>\Omega$, where cooling and squeezing effects are expected to be stronger.

The ideas presented here can be used to investigate the physical behavior of an impurity in a BEC by open quantum systems techniques. 
In this framework the impurity plays the role of the Brownian particle, while the set of the BEC Bogoliubov modes represents the environment. 
The linear QBM provides a useful tool to study the dynamics of the impurity in a uniform medium, while the QBM with a generic coupling may be used to investigate impurities immersed in an inhomogeneous background, such as the one provided by an harmonic trap.

In conclusion, in Appendix \ref{appendixGauss} we proved that the Gaussian approximation preserves the Lindblad form of a ME, and so it does not yield any HUP violation, regardless of whether it is performed on the equations for the moments or directly on the LME.
We developed this demonstration starting from a LME related to a Lindblad operator which is just quadratic in the creation and annihilation operators, because it is enough to cover the situation analysed in Sec.\ \ref{QuadraticCase}. 
In general, one could extend the proof to LMEs associated to Lindblad operators containing $n^{\rm th}$ powers of creation and annihilation operators.  
This, as far as we know, has never been shown and constitutes an interesting motivation for future projects. Also, a generalization of this proof to LMEs for fermionic systems \cite{Kraus2009} is apparently possible and interesting.

\acknowledgments This work has been funded by a scholarship from the Programa M\'{a}sters d'Excel-l\'{e}ncia of the Fundaci\'{o} Catalunya-La Pedrera, ERC Advanced Grant OSYRIS, EU IP SIQS, EU PRO QUIC, 
EU STREP EQuaM (FP7/2007-2013, No. 323714), Fundaci\'o Cellex, the Spanish MINECO (SEVERO OCHOA GRANT SEV-2015-0522 and FOQUS FIS2013-46768), the Generalitat de Catalunya (SGR 874).
P.M. is funded by a ``Ram\'on y Cajal" fellowship.  S.H.L. and J.W. were partially supported by the NSF grant MS 131271.  They are grateful to L. Torner and ICFO for hospitality in the Summer of 2015.

\appendix
\section{Heisenberg Uncertaintly Principle for density operators}\label{appendixHUP}
The purpose of this Appendix is to present a self-contained derivation of the Heisenberg uncertainty principle for density operators. 
We start from the pure state case.  Consider an arbitrary state $|\psi\rangle$ and observables $\hat{A}$ and $\hat{B}$.  Denoting by $\langle \hat{A} \rangle$ the mean of the observable $\hat{A}$ in the state $\left|\psi\right>$,
\beq
\ave{\hat{A}} = \bra{\psi}\hat{A}\ket{\psi},
\eeq
for the variance of  $\hat{A}$ we have
\beq
\sigma_A^2 = \left<\psi\left|\left(\hat{A} - \ave{\hat{A}}\right)^2\right|\psi\right>,
\eeq
and similarly for $\hat{B}$.  For future reference, let us also note that for any real number $a$, 
\beq\label{property}
 \left<\psi\left|\left(\hat{A} - a\right)^2\right|\psi\right> \geq \sigma_A^2.
\eeq
The claim we want to prove is
\beq
\sigma_A^2\sigma_B^2 \geq  {1 \over 4}\left<{[\hat{A}, \hat{B}] \over i}\right>^2,
\eeq
where the right-hand side contains the mean value of the observable ${[\hat{A}, \hat{B}] / i}$ in the state $\left|\psi\right>$.
Introducing the vectors
\beq
\left|f\right>= \left|(\hat{A}-\ave{\hat{A}})\psi\right> \quad \hbox{and} \quad \left|g\right>= \left|(\hat{B} - \ave{\hat{B}})\psi\right>
\eeq
we have
\beq
\sigma_A^2 \sigma_B^2 = \left<f|f\right>\left<g|g\right> \geq \left|\left<f|g\right>\right|^2
\eeq
applying the Cauchy-Schwarz inequality.  The right-hand side of the last inequality can be rewritten as 
\beq
\left|\left<f|g\right>\right|^2 = \left({\left<f|g\right> + \left<g|f\right> \over 2}\right)^2 + \left({\left<f|g\right> - \left<g|f\right> \over 2i}\right)^2
\eeq
with both terms on the right-hand side nonnegative.  Rewriting the second term as the square of the mean of the observable ${[\hat{A}, \hat{B}] / 2i}$, and leaving the first term out (keeping it would lead to a stronger inequality, called Robertson-Schr\"odinger inequality), we obtain the desired bound
\beq
\sigma_A^2 \sigma_B^2 \geq {1 \over 4}\left<{[\hat{A}, \hat{B}] \over i}\right>^2
\eeq
in the pure state case.  Now, if $\rho = \sum_j p_j \left|\phi_j\right>\left<\phi_j\right|$ is an arbitrary density operator, with $p_j$ non-negative coefficients summing up to 1, the mean of $\hat{A}$ in the state $\rho$ equals
\beq
 \ave{\hat{A}}^{(\rho)} =\text{Tr}\left(\hat \rho \hat{A}\right).
\eeq
For the variance of $\hat{A}$ in the state $\rho$ we have
\beq
\left(\sigma_A^{(\rho)}\right)^2 =\text{Tr}\left[\hat \rho\left(\hat{A} - \ave{ \hat{A}}^{(\rho)}\right)^2\right],
\eeq
and similarly for $\hat{B}$.  We thus have
\begin{align}
\left(\sigma_A^{(\rho)}\right)^2&= \sum_jp_j \left<\phi_j\left|\left(\hat{A} - \ave{\hat{A}}^{(\rho)}\right)^2\right|\phi_j\right>\nonumber\\
&\geq \sum_jp_j\left(\sigma_A^{(\phi_j)}\right)^2
\end{align}
where $\left(\sigma_A^{(\phi_j)}\right)^2$ denotes the variance of $\hat{A}$ in the state $\left|\phi_j\right>$, and in the last step we used inequality Eq.\ \eqref{property}.  Similarly, 
\beq
\left(\sigma_B^{(\rho)}\right)^2 \geq \sum_jp_j\left(\sigma_B^{(\phi_j)}\right)^2
\eeq
By the (discrete version of) the Cauchy-Schwarz inequality (it is crucial that $p_j \geq 0$ here!) we obtain
\beq
\left(\sigma_A^{(\rho)}\right)^2\left(\sigma_B^{(\rho)}\right)^2 \geq \left(\sum_jp_j\sigma_A^{(\phi_j)}\sigma_B^{(\phi_j)}\right)^2
\eeq
which, using the pure-state version of the uncertainty principle, is bounded from below by
\beq
{1 \over 4}\left(\sum_jp_j\left\langle\phi_j\left|{\left[\hat{A},\hat{B}\right] \over i}\right|\phi_j\right\rangle\right)^2 = {1 \over 4}\left(\left<{\left[\hat{A},\hat{B}\right] \over i}\right>^{(\rho)}\right)^2
\eeq
which is the desired mixed-state version of the inequality.  In the last application of the Cauchy-Schwarz inequality, it is crucial that we are dealing with a density operator, so that the eigenvalues $p_j$ are nonnegative.

In the case most important for us, when $\hat{A} = \hat{X}$ is the position operator and $\hat{B} = \hat{P}$ is the momentum operator, the commutator of $\hat{A}$ and $\hat{B}$ is a multiple of identity, $\left[\hat{X},\hat{P}\right] = i\hbar \hat{I}$.  The mean value of ${\left[\hat X,\hat P\right] \over i}$ in any state is thus equal to $\hbar$ and in particular, for the density operators $\hat{\rho}_t$, solving a Lindblad equation we obtain at all times the standard form of the Heisenberg Uncertainty Principle,
\beq
\sigma_X^2\sigma_P^2 \geq {\hbar^2 \over 4}.
\eeq

\section{Gaussian Approximation}\label{appendixGauss}
The purpose of this Appendix is to prove that the Gaussian approximation performed on the LME for Quadratic QBM preserves its Lindblad form. 
The demonstration we are about to present considers a Gaussian approximation carried out directly on the ME, while in Section \ref{sec:StationaryStateQuadratic} it has been done on the equations for the moments. 
As we will show, the two procedures are completely equivalent. \\

\noindent{\bf Theorem}
For a quadratic Lindblad operator: 
\beq\label{LindOpQuadAlternative}
\hat{L}=\tilde{\alpha}\hat{a}^2+\tilde{\beta}(\hat{a}^{\dag})^2+\tilde{\gamma}\hat{a}^{\dag}\hat{a}+\tilde{\delta}\hat{a}+\tilde{\epsilon} \hat{a}^{\dag}+\tilde{\eta}
\eeq
the self-consistent Gaussian approximation preserves the Lindblad form (and thus the positivity of $\hat{\rho}$ and HUP).

The annihilation and creation operators are represented respectively by $\hat{a}$ and $\hat{a}^{\dag}$, while $\tilde{\alpha}, \tilde{\beta}, \tilde{\gamma}, \tilde{\delta}, \tilde{\epsilon},\tilde{\eta}$ are complex parameters. 
It is immediate to prove that the Lindblad operator introduced in Eq.\ (\ref{LindOpQuadQBM}) can be expressed in the form showed in Eq.\ (\ref{LindOpQuadAlternative}).
Note that it is possible to assume $\ave{\hat{a}}=0$, since it just shifts the parameters. \\

\noindent{\bf Lemma 1} The parameter $\tilde{\eta}$ in Eq.\ (\ref{LindOpQuadAlternative}) can be shifted arbitrarily. \\

\noindent {\it Proof:} the core of the proof lies in the fact that any additive constant in the definition of the Lindblad operator can be compensated by a re-definition of the Hamiltonian, namely:
\begin{align}
\der{\hat{\rho}}{t}&=-\frac{i}{\hbar}[\hat{H},\hat{\rho}]+\mathcal{D}_{L+\Delta\tilde\eta}(\hat{\rho})\\ \nonumber
&=-\frac{i}{\hbar}[\hat{H}+\Delta\hat{H}_{\Delta\tilde{\eta}},\hat{\rho}]+\mathcal{D}_{L}(\hat{\rho}),
\end{align}
where:
\begin{equation}
\mathcal{D}_{L}(\hat{\rho})=\hat{L}\hat{\rho}\hat{L}^{\dag} - \hat{L}^{\dag}\hat{L}\hat{\rho}/2 -  \hat{\rho}\hat{L}^{\dag}\hat{L}/2, 
\end{equation}
is the Lindblad dissipator, and:
\begin{equation}
\Delta\hat{H}_{\Delta\tilde{\eta}}=-\frac{i}{2}[(\Delta\tilde\eta) \hat{L}^{\dag}-(\Delta\tilde \eta)^*\hat{L}],
\end{equation}
with $\Delta\tilde{\eta}\in\mathbb{C}$. 

Of course changing of Hamiltonian is allowed, since it just modifies the time dependence of $\hat{a}$ and $\hat{a}^{\dag}$ in the interaction picture.\\

\noindent{\bf Lemma 2} It is possible to perform the factorization:
\beq\label{FactLindOp}
\hat{L}=\hat{d}_1\hat{d}_2
\eeq
with:
\begin{align}\label{d1d2}
&\hat{d}_1=\tilde{A}\hat{a}+\tilde{B}\hat{a}^{\dag}+\tilde{C}\\ \nonumber
&\hat{d}_2=\hat{a}+\tilde{D}\hat{a}^{\dag}+\tilde{E}
\end{align}

\noindent {\it Proof:}
comparing Eqs.\ (\ref{FactLindOp}) and (\ref{LindOpQuadAlternative}), one obtains:
\begin{align}
&\tilde{A}=\tilde{\alpha},\quad \tilde{A}\tilde{D}+\tilde{B}=\tilde{\gamma}, \quad \tilde{A}\tilde{D}+\tilde{C}\tilde{E}=\tilde{\eta}\\ \nonumber
&\tilde{B}\tilde{D}=\tilde{\beta},\quad
\tilde{A}\tilde{E}+\tilde{C}=\tilde{\delta},\quad \tilde{B}\tilde{E}+\tilde{C}\tilde{D}=\tilde{\delta},
\end{align}
so that
\begin{align}
\tilde{D}=\tilde{\beta}/\tilde{B},\quad \tilde{\alpha}\tilde{\beta}/\tilde{B}+\tilde{B}=\tilde{\gamma}
\end{align}
provide in general two solutions $\tilde{B}_1$ and $\tilde{B}_2$ for $\tilde{B}$, and
\begin{equation}
\tilde{\alpha}\tilde{E}+\tilde{C}=\tilde{\delta},\quad\tilde{B}\tilde{E}+(\tilde{\beta}/\tilde{B})\tilde{C}=\tilde{\eta}.
\end{equation}
 
If we can solve these linear equations for $\tilde{E}$ and $\tilde{C}$, we 
may plug the  solution into
$\tilde{\alpha}\tilde{D}+\tilde{C}\tilde{E}=\tilde{\eta}$, and  adjust $\eta$ adequately (which we can do according to Lemma 1). 

It is easy to check that the two equations for $\tilde{E}$ and $\tilde{C}$ cannot be solved if $\tilde{B}_1=\tilde{B}_2=0$, which implies $\tilde{\gamma} =0$
and $\tilde{\alpha}\tilde{\beta}=0$, i.e. the non-generic case $\hat{L}=\tilde{\alpha} \hat{a}^2+\tilde{\delta}\hat{a}+\tilde{\epsilon}\hat{a}^{\dag}+\tilde{\eta}$, and the related one
with $\tilde{\alpha}=0$ , $\tilde{\beta}\ne 0$. 
The case $\tilde{\alpha}=\tilde{\beta}=0$ is trivial, as it corresponds to linear Lindblad operator: for such a case, the Gaussian approximation is not needed, since there exists an exact solution of Gaussian form. 

Now we prove the Theorem in the generic case: 

\noindent {\it Proof of the Theorem:} We look to the Lindblad dissipator related to the factorized Lindblad operator in Eq. (\ref{FactLindOp}):
\begin{equation}\label{ExactSuper}
\mathcal{D}_{L}(\hat{\rho})=\hat{d}_1\hat{d}_2\hat{\rho} \hat{d}_2^{\dag}\hat{d}_1^{\dag}-\frac{1}{2}\{\hat{d}_2^{\dag}\hat{d}_1^{\dag}\hat{d}_1\hat{d}_2,\hat{\rho}\}
\end{equation}
In the Gaussian approximation, one replaces pairs of operators by their mean values.
``Anomalous" terms
generate contributions that may be reabsorbed in the Hamiltonian, such as
\begin{align}\label{apprTermAn1}
&\ave{\hat{d}_1\hat{d}_2}\left[\hat{\rho} \hat{d}_2^{\dag}\hat{d}_1^{\dag}- \frac{1}{2}\{\hat{d}_2^{\dag}\hat{d}_1^{\dag},\hat{\rho}\}\right]=-\frac{1}{2}\ave{\hat{d}_1\hat{d}_2}[\hat{d}_2^{\dag}\hat{d}_1^{\dag},\hat{\rho}]
\end{align}
and:
\begin{align}\label{apprTermAn2}
\ave{\hat{d}^\dagger_2\hat{d}^\dagger_1}\left[\hat{d}_1\hat{d}_2\hat{\rho}-\frac{1}{2}\{\hat{d}_1\hat{d}_2,\hat{\rho}\}\right]=\frac{1}{2}\ave{\hat{d}^\dagger_2\hat{d}^\dagger_1}[\hat{d}_1\hat{d}_2,\hat{\rho}].
\end{align}
The non-trivial terms are:
\begin{align}\label{apprTermJump}
&\ave{\hat{d}^\dag_2 \hat{d}_1}\hat{d}_2\hat{\rho} \hat{d}_1^{\dag}+
\ave{\hat{d}^\dag_1 \hat{d}_1}\hat{d}_2\hat{\rho} \hat{d}_2^{\dag}\\ \nonumber
+&\ave{\hat{d}^\dag_2 \hat{d}_2}\hat{d}_1\hat{\rho} \hat{d}_1^{\dag}
+\ave{\hat{d}^\dag_1 \hat{d}_2}\hat{d}_1\hat{\rho} \hat{d}_2^{\dag}\\ \nonumber
-&\left\{\left(\ave{\hat{d}_2^{\dag}\hat{d}_1} \hat{d}_1^{\dag}\hat{d}_2 +
\ave{\hat{d}_2^{\dag}\hat{d}_2}\hat{d}_1^{\dag}\hat{d}_1\right),\frac{\hat{\rho}}{2}\right\}\\ \nonumber
-&\left\{\left(
\ave{\hat{d}_1^{\dag}\hat{d}_1}\hat{d}_2^{\dag}\hat{d}_2 +
\ave{\hat{d}_1^{\dag}\hat{d}_2}\hat{d}_2^{\dag}\hat{d}_1\right),\frac{\hat{\rho}}{2}\right\}
\end{align}

The resulting ME has a dissipator of the form:
\begin{equation}\label{FinalLME}
\mathcal{D}_L(\hat{\rho})=\sum_{i,j=1,2} \tilde{\Gamma}_{ij}\left(\hat{d}_i\hat{\rho}\hat{d}_j^{\dag}-\frac{1}{2}\{\hat{d}_j^{\dag} \hat{d}_i,\hat{\rho}\}\right),
\end{equation}
where $\tilde{\Gamma}_{ij}=\ave{\hat{d}^{\dag}_{j'}\hat{d}_{i'}}$, where $1'=2$ and $2'=1$.
This matrix is evidently positive definite, as follows from the Schwartz inequality, so that the dissipator is again of Lindblad form.

Note that the generalization to many oscillators, many Lindblad operators is straightforward. 
Note also that the non-generic case is simple to treat. It requires, however, a direct calculation. The quartic Lindblad term in this case is treated as above, while the quadratic one does not need to be touched, since it already describes a Gaussian quantum process. The third order term on the other hand partially vanishes and partially gives contributions to the Hamiltonian in the Gaussian approximation. 

The remaining question is whether the approximation that we perform on the level of the ME is the same as the Gaussian de-correlation we performed according to the Wick's theorem prescription at the level of the equations for the moments in Sec.\ \ref{sec:StationaryStateQuadratic}. To illustrate this, we consider an arbitrary operator $\hat{O}$ and we derive the dynamical equations for its average value starting by the ME induced by the superoperator in Eq. (\ref{ExactSuper}).

The dynamical equation for the average value of an operator $\hat{O}$ presents the following form:
\begin{equation}\label{eqDinO}
\der{\ave{\hat{O}}}{t}=h^{u}_{O}+h^{(1)}_{O}-\frac{1}{2}\left(h^{(2)}_{O}+h^{(3)}_{O}\right)
\end{equation}
in which:
\begin{align}\label{avVal}
&h^{u}_{O}= -\frac{i}{\hbar}{\rm Tr}\left(\hat{O}\left[\hat{H},\hat{\rho}\right]\right)\\\nonumber
&h^{(1)}_{O}={\rm Tr}(\hat{O}\hat{d}_1 \hat{d}_2 \hat{\rho}  d^\dag_2 \hat{d}_1^{\dag}) =\langle{ \hat{d}_2^{\dag} \hat{d}^\dag_1 \hat{O} \hat{d}_1 \hat{d}_2}\rangle\\
&h^{(2)}_{O}={\rm Tr}(\hat{O} \hat{d}^\dag_2 \hat{d}_1^{\dag}\hat{d}_1 \hat{d}_2 \hat{\rho} ) =\langle{ \hat{O} \hat{d}_2^{\dag} \hat{d}^\dag_1 \hat{d}_1 \hat{d}_2}\rangle\nonumber\\
&h^{(3)}_{O}={\rm Tr}(\hat{O}\hat{\rho}\hat{d}^\dag_2 \hat{d}_1^{\dag}\hat{d}_1 \hat{d}_2 ) =\langle{\hat{d}_2^{\dag} \hat{d}^\dag_1 \hat{d}_1 \hat{d}_2\hat{O} }\rangle\nonumber.
\end{align}
Performing the Gaussian approximation at the level of the equation for the moments means to carry out such an approximation on the average values in Eqs.\ (\ref{avVal}),
\begin{eqnarray}\label{eqMot1}
&h^{(1)}_{O}={\rm Tr}(\hat{O}\hat{d}_1 \hat{d}_2 \hat{\rho}  d^\dag_2 \hat{d}_1^{\dag}) =\langle{ \hat{d}_2^{\dag} \hat{d}^\dag_1 \hat{O} \hat{d}_1 \hat{d}_2}\rangle\\  \nonumber
\simeq& \langle{\hat{d}^\dag_2 \hat{d}_1^{\dag}}\rangle\langle{\hat{O}\hat{d}_1 \hat{d}_2}\rangle + \langle{d^\dag_2 \hat{d}_1^{\dag}\hat{O}}\rangle\langle{\hat{d}_1 \hat{d}_2}\rangle  - \langle{\hat{d}^\dag_2 \hat{d}_1^{\dag}\rangle\langle\hat{O}}\rangle\langle{\hat{d}_1 \hat{d}_2}\rangle\\ \nonumber
+& \langle{\hat{d}^\dag_2 \hat{d}_1}\rangle\langle{\hat{d}_1^{\dag}\hat{O} \hat{d}_2}\rangle + \langle{\hat{d}^\dag_2 \hat{O} \hat{d}_1}\rangle\langle{\hat{d}_1^{\dag} \hat{d}_2}\rangle  - \langle{\hat{d}^\dag_2 \hat{d}_1\rangle\langle \hat{O}}\rangle\langle{\hat{d}_1^{\dag} \hat{d}_2}\rangle\\ \nonumber
+& \langle{\hat{d}^\dag_2 \hat{d}_2}\rangle\langle{\hat{d}_1^{\dag} \hat{O} \hat{d}_1}\rangle + \langle{\hat{d}^\dag_2 \hat{O} \hat{d}_2}\rangle\langle{\hat{d}_1^{\dag} \hat{d}_1}\rangle  - \langle{\hat{d}^\dag_2 \hat{d}_2\rangle\langle \hat{O}}\rangle\langle{\hat{d}_1^{\dag} \hat{d}_1}\rangle,\nonumber
\end{eqnarray}
\begin{eqnarray}\label{eqMot2}
&h^{(2)}_{O}={\rm Tr}(\hat{O} \hat{d}^\dag_2 \hat{d}_1^{\dag}\hat{d}_1 \hat{d}_2 \hat{\rho} ) =\langle{ \hat{O} \hat{d}_2^{\dag} \hat{d}^\dag_1 \hat{d}_1 \hat{d}_2}\rangle\\  \nonumber
\simeq& \langle{\hat{d}^\dag_2 \hat{d}_1^{\dag}}\rangle\langle{\hat{O}\hat{d}_1 \hat{d}_2}\rangle + \langle{\hat{O} \hat{d}^\dag_2 \hat{d}_1^{\dag}}\rangle\langle{\hat{d}_1 \hat{d}_2}\rangle  - \langle{\hat{d}^\dag_2 \hat{d}_1^{\dag}\rangle\langle \hat{O}}\rangle\langle{\hat{d}_1 \hat{d}_2}\rangle\\ \nonumber
+& \langle{\hat{d}^\dag_2 \hat{d}_1}\rangle\langle{\hat{O} \hat{d}_1^{\dag}  \hat{d}_2}\rangle + \langle{\hat{O} \hat{d}^\dag_2  \hat{d}_1}\rangle\langle{\hat{d}_1^{\dag} \hat{d}_2}\rangle  - \langle{\hat{d}^\dag_2 \hat{d}_1\rangle\langle \hat{O}}\rangle\langle{\hat{d}_1^{\dag} \hat{d}_2}\rangle\\ \nonumber
+& \langle{\hat{d}^\dag_2 \hat{d}_2}\rangle\langle{\hat{O} \hat{d}_1^{\dag}  \hat{d}_1}\rangle + \langle{\hat{O} \hat{d}^\dag_2 \hat{d}_2}\rangle\langle{\hat{d}_1^{\dag} \hat{d}_1}\rangle  - \langle{d^\dag_2 \hat{d}_2\rangle\langle \hat{O}}\rangle\langle{\hat{d}_1^{\dag} \hat{d}_1}\rangle,\nonumber
\end{eqnarray} 
\begin{eqnarray}\label{eqMot3}
&h^{(3)}_{O}={\rm Tr}(\hat{O} \hat{\rho} \hat{d}^\dag_2 \hat{d}_1^{\dag}\hat{d}_1 \hat{d}_2) =\langle{  \hat{d}_2^{\dag} \hat{d}^\dag_1 \hat{d}_1 \hat{d}_2 \hat{O}}\rangle\\  \nonumber
\simeq& \langle{\hat{d}^\dag_2 \hat{d}_1^{\dag}}\rangle\langle{\hat{d}_1 \hat{d}_2 \hat{O}}\rangle + \langle{ \hat{d}^\dag_2 \hat{d}_1^{\dag}\hat{O}}\rangle\langle{\hat{d}_1 \hat{d}_2}\rangle  - \langle{\hat{d}^\dag_2 \hat{d}_1^{\dag}\rangle\langle \hat{O}}\rangle\langle{\hat{d}_1 \hat{d}_2}\rangle\\ \nonumber
+& \langle{\hat{d}^\dag_2 \hat{d}_1}\rangle\langle{ \hat{d}_1^{\dag}  \hat{d}_2 \hat{O}}\rangle + \langle{\hat{d}^\dag_2  \hat{d}_1 \hat{O}}\rangle\langle{\hat{d}_1^{\dag} \hat{d}_2}\rangle  - \langle{\hat{d}^\dag_2 \hat{d}_1\rangle\langle \hat{O}}\rangle\langle{\hat{d}_1^{\dag} \hat{d}_2}\rangle\\ \nonumber
+& \langle{\hat{d}^\dag_2 \hat{d}_2}\rangle\langle{\hat{d}_1^{\dag}  \hat{d}_1 \hat{O}}\rangle + \langle{\hat{d}^\dag_2 \hat{d}_2 \hat{O}}\rangle\langle{\hat{d}_1^{\dag} \hat{d}_1}\rangle  - \langle{d^\dag_2 \hat{d}_2\rangle\langle \hat{O}}\rangle\langle{\hat{d}_1^{\dag} \hat{d}_1}\rangle. \nonumber
\end{eqnarray}

It is now tedious but easy to check that replacing the expressions in Eq.\ (\ref{eqMot1}-\ref{eqMot3}) in Eq.\ (\ref{eqDinO}) we get the dynamical equations generated by the terms in Eqs.\ (\ref{apprTermAn1}-\ref{apprTermJump}), obtained by performing the Gaussian approximation on the ME related to a dissipator in Eq. (\ref{ExactSuper}). This proves that performing the Gaussian approximation at the level of the ME is equivalent to doing it at the level of the equations for the moments of an observable.  
Note that the equations resulting by this approximation will always admit a Gaussian solution, although it is not guaranteed that the latter is stationary.

The demonstration we developed holds for Lindblad operators which are quadratic in the creation and annihilation operators. This case covers the situation studied in Sec. (\ref{QuadraticCase}), but it is not the most general one. 
In fact, one could consider also LMEs with Lindblad operators containing higher powers of creation and annihilation operators. 
Extending the proof we presented to this general case is an interesting perspective that we reserve for future works.

\bibliography{LindbladQBM}
\end{document}